\documentclass[aps,prd,twocolumn,nofootinbib,nobibnotes]{revtex4-1}
\pdfoutput=1

\usepackage[usenames,dvipsnames]{xcolor}
\usepackage{hyperref}
\hypersetup{
	colorlinks=true,
	urlcolor=Maroon,
	linkcolor=Maroon,
	citecolor=Maroon,
	bookmarks=true,
	pdftitle={Scattering Amplitudes from Intersection Theory},
	pdfauthor={Sebastian Mizera},
	pdfdisplaydoctitle=true,
	pdfstartview=FitH
}

\usepackage{amsmath}
\usepackage{amsfonts}
\usepackage{mathtools}
\usepackage{microtype}
\usepackage{tikz}
\usepackage{bm}
\usepackage[utf8]{inputenc}
\usepackage{blkarray}
\usepackage{bigstrut}
\usepackage{nccmath}
\usepackage{enumitem}
\usepackage{braket}
\usepackage{dsfont}
\usepackage[english]{babel}

\newcommand\topstrut[1][1.2ex]{\setlength\bigstrutjot{#1}{\bigstrut[t]}}
\newcommand\botstrut[1][0.9ex]{\setlength\bigstrutjot{#1}{\bigstrut[b]}}

\allowdisplaybreaks

\DeclareMathOperator{\Gr}{Gr}
\DeclareMathOperator{\Res}{Res}
\DeclareMathOperator{\Pf}{Pf}
\DeclareMathOperator{\SL}{SL}

\def \be {\begin{equation}}
\def \ee {\end{equation}}
\def \nn {\nonumber}
\def \la {\left<}
\def \ra {\right>}
\def \w  {\wedge}
\def \PT {\mathsf{PT}}
\def \F  {\mathsf{F}}
\def \C  {\mathsf{C}}

\definecolor{color2}{rgb}{0.368417, 0.506779, 0.709798}
\definecolor{color3}{rgb}{0.880722, 0.611041, 0.142051}
\definecolor{color5}{rgb}{0.560181, 0.691569, 0.194885}
\definecolor{color1}{rgb}{0.922526, 0.385626, 0.209179}
\definecolor{color6}{rgb}{0.528488, 0.470624, 0.701351}
\definecolor{color4}{rgb}{0.772079, 0.431554, 0.102387}
\definecolor{color8}{rgb}{0.5, 0.5, 0.5}
\definecolor{color7}{rgb}{0.6, 0.6, 0.6}

\def \f[#1] {{\color{color#1}\bm{f_#1}}}
\def \a[#1] {{\color{color#1}\bm{\alpha_#1}}}
\def \col[#1]#2 {{\color{color#1}\bm{#2}}}
\def \cov[#1]   {\genfrac{}{}{0pt}{0}{}{\medmath{#1}}}
\def \contr[#1] {\genfrac{}{}{0pt}{0}{\medmath{#1}}{}}

\begin{document}

\title{Scattering Amplitudes from Intersection Theory}
\author{Sebastian Mizera\vspace{.3em}}
\email{smizera@pitp.ca}
\affiliation{Perimeter Institute for Theoretical Physics, Waterloo, ON N2L 2Y5, Canada}
\affiliation{Department of Physics \& Astronomy, University of Waterloo, Waterloo, ON N2L 3G1, Canada}

\begin{abstract}\noindent
We use Picard--Lefschetz theory to prove a new formula for intersection numbers of twisted cocycles associated to a given arrangement of hyperplanes. In a special case when this arrangement produces the moduli space of punctured Riemann spheres, intersection numbers become tree-level scattering amplitudes of quantum field theories in the Cachazo--He--Yuan formulation.
\end{abstract}

\maketitle

\section{Introduction}

\noindent\textsc{Over the last years}, study of scattering amplitudes revealed many unexpected connections to geometric structures \cite{Witten:2003nn,ArkaniHamed:2012nw,Arkani-Hamed:2013jha,Atiyah:2017erd}, allowing us to understand physical properties of quantum field theories---such as locality or unitarity---from a different perspective. At the same time, they equip us with new mathematical tools that vastly simplify practical calculations. In this work we unravel another connection to a branch of mathematics called \emph{intersection theory} \cite{eisenbud20163264,cho1995,aomoto2011theory,yoshida2013hypergeometric}.

It has recently transpired that intersection theory plays an important role in string theory amplitudes, where in particular it provides a geometric interpretation of the Kawai--Lewellen--Tye (KLT) relations between open and closed string amplitudes, or---in the field-theory limit---Yang--Mills and Einstein gravity amplitudes \cite{Kawai:1985xq,Mizera:2017cqs}. Here, we show that analogous structures appear directly in scattering amplitudes of ordinary quantum field theories. We find that they can be understood as intersection numbers of the so-called \emph{twisted cocycles} \cite{cho1995,aomoto2011theory,yoshida2013hypergeometric}, which are certain families of differential forms.

It is instructive to start with an explicit example straight away. Let us consider $\mathbb{CP}^2$ with inhomogeneous coordinates $(x,y)$, dissected by six hyperplanes defined through linear equations $\{f_i \!=\! 0\}$. We can easily visualize the real section of this space with a concrete choice of hyperplanes, for instance:\vspace{-0.5em}
\be
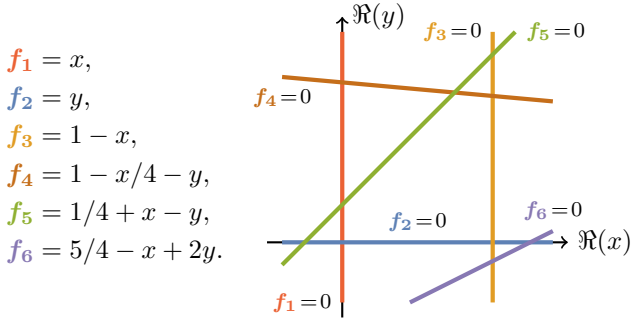
\begin{tabular*}{\columnwidth}{c @{\extracolsep{\fill}} c}
${\begin{aligned}
	&\f[1] = x,\\
	&\f[2] = y,\\
	&\f[3] = 1-x,\\
	&\f[4] = 1-x/4-y,\\
	&\f[5] = 1/4+x-y,\\
	&\f[6] = 5/4-x+2y.
\end{aligned}}$
&\begin{tikzpicture}[scale=2,baseline=3em]
\draw[thick,->] (-0.5,0) -- (1.5,0) node[right] {$\Re(x)$};
\draw[thick,->] (0,-0.5) -- (0,1.5) node[right] {$\Re(y)$};
\draw[ultra thick,color1] (0,-0.4) node[left] {$\medmath{\f[1] \color{black}\!=\!0}$} -- (0,1.4);
\draw[ultra thick,color2] (-0.4,0) -- node[above] {$\medmath{\f[2] \color{black}\!=\!0}$} (1.4,0);
\draw[ultra thick,color3] (1,-0.4) -- (1,1.4) node[left] {$\medmath{\f[3] \color{black}\!=\!0}$};
\draw[ultra thick,color4] (-0.4,1+0.4/4) node[below] {$\medmath{\f[4] \color{black}\!=\!0}$} -- (1.4,1-1/4/4) ;
\draw[ultra thick,color5]  (-0.4,1/4-0.4) -- (1.4-1/4,1.4) node[right] {$\medmath{\f[5] \color{black}\!=\!0}$};
\draw[ultra thick,color6]  (5/4-2*0.4,-0.4) -- (1.4,-5/8+1.4/2) node[above] {$\medmath{\f[6] \color{black}\!=\!0}$};
\end{tikzpicture}\nn\\
\end{tabular*}
\ee
The space of our interest is the original manifold with these hyperplanes removed:\vspace{-0.5em}
\be
X = \mathbb{CP}^2 \setminus \bigcup_{i=1}^{6} \{ f_i = 0 \}.
\ee
Associated to it, we can define a differential 1-form $\omega$, called the \emph{twist}, with logarithmic singularities along $f$'s:\vspace{-0.5em}
\begin{align}\label{twist}
\omega =& \sum_{i=1}^{6} \alpha_i\, d\log f_i\\
=& \underbrace{\left(\medmath{\frac{\a[1] }{x} \!+\! \frac{\a[3] }{-1\!+\!x} \!+\! \frac{\a[4] }{-4\!+\!x\!+\!4y} \!+\! \frac{\a[5] }{1/4\!+\!x\!-\!y} \!+\! \frac{\a[6] }{-5/4\!+\!x\!-\!2y}} \right)}_{\medmath{\omega_x}} dx\nn\\[-.5em]
&\;\,+ \underbrace{\left(\medmath{\frac{\a[2] }{y} + \frac{\a[4] }{-1\!+\!x/4\!+\!y} + \frac{\a[5] }{-1/4\!-\!x\!+\!y} + \frac{\a[6] }{5/8\!-\!x/2\!+\!y}} \right)}_{\medmath{\omega_y}} dy,\nn
\end{align}
where $\alpha$'s are constant coefficients adding up to zero. The twist $1$-form fully characterizes the space $X$.

On this space we can introduce two differential forms, $\varphi_L$ and $\varphi_R$. We choose them in such a way that they have logarithmic singularities on three of the hyperplanes defined above. For instance, we can take:
\begin{align}
\varphi_L =& d\log \frac{\f[2] }{\f[3] } \w d\log \frac{\f[3] }{\f[5] } = \frac{5\, dx \w dy}{y(1-x)(1+4x-4y)},\\
\varphi_R =& d\log \frac{\f[2] }{\f[3] } \w d\log \frac{\f[3] }{\f[6] } = \frac{dx \w dy}{y(1-x)(5-4x+8y)}.
\end{align}
These objects are examples of twisted cocycles, which roughly speaking are differential forms on $X$ defined up to equivalence classes $\varphi \sim \varphi + \omega \w \xi$ for any $d\!\log$ form $\xi$. One can define an invariant pairing called the \emph{intersection number}\footnote{Intersection numbers are normally defined for \emph{cycles}. It is conventional \cite{cho1995,aomoto2011theory,yoshida2013hypergeometric} to use the same name for a pairing of cocycles, even though its geometrical interpretation is different. For \emph{twisted} cohomologies \cite{aomoto2011theory}, intersection numbers are in general not integers.}. Its standard definition \cite{cho1995,aomoto2011theory} reads:
\be\label{intro-intersection-number-definition-old}
\la \varphi_L, \varphi_R \ra_\omega = \frac{1}{(2\pi i)^2}\int_X \iota_\omega (\varphi_L) \w \varphi_R,
\ee
where the map $\iota_\omega$ turns $\varphi_L$ into its compactly supported version, i.e., one that vanishes in a small neighbourhood of the hyperplanes $\{f_i \!=\! 0\}$. Note that the integrand would vanish if it was not for this map. Here we also remark that $\langle \varphi_L |$ and $| \varphi_R \rangle$ belong to different cohomologies, as will be discussed in the following section. A result of this calculation reveals a combinatorial formula \cite{matsumoto1998}:
\pagebreak

\be\label{intro-intersection-number-old}
\la \varphi_L, \varphi_R \ra_\omega = \;\pm\!\!\!\!\!\!\!\! \sum_{\{f_i, f_j\} \in L, R} \frac{1}{\alpha_i\, \alpha_j} = \frac{1}{\a[2] \, \a[3] }.
\ee
We review the meaning of the map $\iota_\omega$ in Appendix~\ref{appendix-a}, which also illustrates how factors of $\alpha_i$ arise in the denominators. The above result is a sum over all intersection vertices of the hyperplanes that are associated to both $\varphi_L$ and $\varphi_R$. In our example, we have $L=(\f[2] ,\f[3] ,\f[5] )$ and $R=(\f[2] ,\f[3] ,\f[6] )$, which intersect at a single point \begin{tikzpicture}[scale=2,baseline=-0.3em]
\draw[ultra thick,color2] (-0.1,0) -- (0.1,0);
\draw[ultra thick,color3] (0,-0.1) -- (0,0.1);
\end{tikzpicture}, and we inserted the correct sign \cite{matsumoto1998}. An important feature of the above formula is that it is completely independent of the precise positions of the hyperplanes, as long as their arrangement is generic, i.e., no three $f$'s intersect at a single point.

In this work we propose an alternative formula for computing intersection numbers as an integral localizing on the points $(x^\ast, y^\ast)$ at which $\omega$ vanishes:
\begin{align}\label{intro-intersection-number-definition-new}
\hspace{-4.3em}\la \varphi_L, \varphi_R \ra_\omega &= \int dx\, dy\, \delta(\omega_x) \delta(\omega_y)\, \widehat{\varphi}_L\, \widehat{\varphi}_R\\
&= \!\!\sum_{(x^\ast, y^\ast)} \!\! {\det}^{-1} \begin{bmatrix} \tfrac{\partial \omega_x}{\partial x} & \tfrac{\partial \omega_x}{\partial y} \\ \tfrac{\partial \omega_y}{\partial x} & \tfrac{\partial \omega_y}{\partial y} \end{bmatrix} \widehat{\varphi}_L\, \widehat{\varphi}_R \,\Bigg|_{(x,y) = (x^\ast, y^\ast)}\hspace{-4.8em},\nn
\end{align}
where we used $\varphi = \widehat{\varphi}\, dx \w dy$. Here, delta functions should be understood as multi-dimensional residue prescriptions around the zeros of $\omega$. Remarkably, this formula evaluates to the rational function of $\alpha$'s \eqref{intro-intersection-number-old} for any choice of $\varphi_L$ and $\varphi_R$, and does so in a highly non-trivial manner.

Readers familiar with scattering amplitudes literature will notice a resemblance of \eqref{intro-intersection-number-definition-new} to the Cachazo--He--Yuan (CHY) formulae \cite{Cachazo:2013hca,Cachazo:2013iea}. This is not a coincidence. In fact, CHY formalism uses a particular, singular, arrangement of hyperplanes, for example:
\be
\{x=0\} \cup \{y=0\} \cup \{1-x=0\} \cup \{1-y=0\} \cup \{x-y=0\}\nn
\ee
with the last hyperplane at infinity, such that the resulting space $X$ is the moduli space of punctured Riemann spheres, in this case $X=\mathcal{M}_{0,5}$. The equation \eqref{intro-intersection-number-old} can no longer be used directly, as the arrangement of hyperplanes is not generic. Nevertheless, the new formula \eqref{intro-intersection-number-definition-new} is still valid. Let us see how this comes about.

We can organize the coefficients of a particular arrangement into a matrix:
\be
C =\;
\begin{blockarray}{ccccccc}
	\f[1] & \f[2] & \f[3] & \f[4] & \f[5] & \f[6] & \\
\begin{block}{[cccccc]c}
	0 & 0 & 1 & 1 & \tfrac{\varepsilon}{4} & \tfrac{4+\varepsilon}{4} & \;1 \topstrut\\
	1 & 0 & -1 & -\tfrac{\varepsilon}{4} & 1 & -\varepsilon & \;x\\
	0 & 1 & 0 & -1 & -1 & 2\varepsilon & \;y \botstrut\\
\end{block}
\end{blockarray}\;,\vspace{-1.5em}
\ee
giving $f_i = c_{1i} + c_{2i}x + c_{3i}y$. We set it up such that $\varepsilon=1$ yields the original arrangement, which deforms into the singular one as $\varepsilon \to 0$. A sign of singularity is that several $ 3\times 3$ minors of $C$ vanish in this limit. Keeping parameters $\alpha$ constant, for example with $\alpha_i > 0$ for $i=1,2,\ldots,5$, the hyperplanes and zeros of $\omega$ behave as follows:
\pagebreak

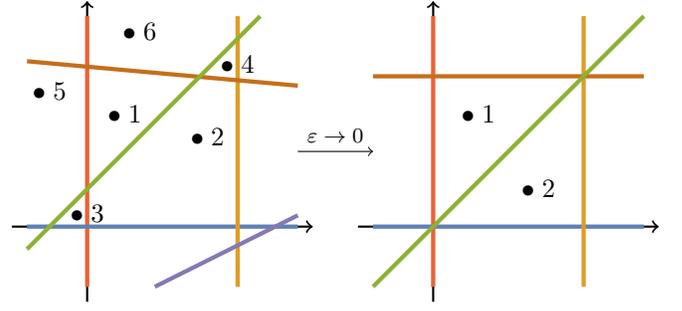
\be
\begin{tikzpicture}[scale=2,baseline=3em]
\draw[thick,->] (-0.5,0) -- (1.5,0);
\draw[thick,->] (0,-0.5) -- (0,1.5);
\draw[ultra thick,color1] (0,-0.4) -- (0,1.4);
\draw[ultra thick,color2] (-0.4,0) -- (1.4,0);
\draw[ultra thick,color3] (1,-0.4) -- (1,1.4);
\draw[ultra thick,color4] (-0.4,1+0.4/4) -- (1.4,1-1/4/4) ;
\draw[ultra thick,color5]  (-0.4,1/4-0.4) -- (1.4-1/4,1.4);
\draw[ultra thick,color6]  (5/4-2*0.4,-0.4) -- (1.4,-5/8+1.4/2);
\node at (0.25,0.75) {\textbullet\;$1$};
\node at (0.8,0.6) {\textbullet\;$2$};
\node at (0.00,0.09) {\textbullet\;$3$};
\node at (1.00,1.08) {\textbullet\;$4$};
\node at (-0.25,0.9) {\textbullet\;$5$};
\node at (0.35,1.3) {\textbullet\;$6$};

\newcommand\shift{2.3}
\draw[->] (1.4,0.5) -- node[above] {\footnotesize$\varepsilon \to 0$} (\shift-0.4,0.5);

\draw[thick,->] (\shift-0.5,0) -- (\shift+1.5,0);
\draw[thick,->] (\shift+0,-0.5) -- (\shift+0,1.5);
\draw[ultra thick,color1] (\shift+0,-0.4) -- (\shift+0,1.4);
\draw[ultra thick,color2] (\shift-0.4,0) -- (\shift+1.4,0);
\draw[ultra thick,color3] (\shift+1,-0.4) -- (\shift+1,1.4);
\draw[ultra thick,color4] (\shift-0.4,1) -- (\shift+1.4,1) ;
\draw[ultra thick,color5]  (\shift-0.4,-0.4) -- (\shift+1.4,1.4);
\node at (\shift+0.3,0.75) {\textbullet\;$1$};
\node at (\shift+0.7,0.25) {\textbullet\;$2$};
\end{tikzpicture}\nn
\ee
The hyperplane $\{\f[6] \!=\! 0\}$ moved to infinity. Out of the six zeros of $\omega$, only $1$ and $2$ survive at finite positions. Both $3$ and $4$ get trapped between three hyperplanes and eventually cease to be zeros of $\omega$ since $(0,0)$ and $(1,1)$ are not a part of the manifold $X$. Similarly, $5$ and $6$ shoot off to infinity. This can be easily verified from the explicit form of the twist in the strict $\varepsilon \to 0 $ limit:
\be
\widetilde{\omega} =\! \left(\medmath{\frac{ \col[1]s_{12} }{x} \!+\! \frac{ \col[3]s_{24} }{x\!-\!1} \!+\! \frac{ \col[5]s_{23} }{x\!-\!y}} \right)\! dx +\! \left(\medmath{ \frac{\col[2]s_{13} }{y} \!+\! \frac{ \col[4]s_{34} }{y\!-\!1} \!+\! \frac{ \col[5]s_{23} }{y\!-\!x}} \right)\! dy,
\ee
where we made an identification of $\alpha$'s with specific Mandelstam invariants, $s_{ab} = (k_a + k_b)^2$ involving ingoing lightlike momenta $k_a$. Note that it preserves the condition $\sum_{i=1}^{6} \alpha_i = 0$ due to momentum conservation. Using a pair of twisted cocycles, for example:\vspace{-0.5em}
\begin{align}
\widetilde{\varphi}_L =& d\log \frac{\f[1] }{\f[5] } \w d\log \frac{\f[5] }{\f[4] } = -\frac{dx \w dy}{x(x-y)(y-1)},\\
\widetilde{\varphi}_R =& d\log \frac{\f[2] }{\f[5] } \w d\log \frac{\f[5] }{\f[3] } = \frac{dx \w dy}{y(y-x)(x-1)},
\end{align}
we can evaluate their intersection number at this singular arrangement via \eqref{intro-intersection-number-definition-new}, giving:
\be
\left< \widetilde{\varphi}_L, \widetilde{\varphi}_R \right>_{\widetilde{\omega}} = \frac{1}{ \col[5]s_{23} } \left( \frac{1}{ \col[1]s_{12} \!+\! \col[2]s_{13} \!+\! \col[5]s_{23} } + \frac{1}{ \col[3]s_{24} \!+\! \col[4]s_{34} \!+\! \col[5]s_{23} } \right),\nn
\ee
which is indeed an example of a bi-adjoint scalar partial amplitude \cite{BjerrumBohr:2012mg,Cachazo:2013iea}. The limit $\varepsilon \to 0$ needs to be taken \emph{before} performing the integral \eqref{intro-intersection-number-definition-new}. Evaluating it at finite $\varepsilon$ yields identically zero in agreement with \eqref{intro-intersection-number-old}, since $L=(\f[1] ,\f[5] ,\f[4] )$ and $R=(\f[2] ,\f[5] ,\f[3] )$ have no intersection points in a generic arrangement. In general, intersection numbers change discontinuously depending on the topology, but not geometry, of the arrangement.

With this example we illustrated how the new prescription \eqref{intro-intersection-number-definition-new} provides a way of calculating intersection numbers even at singular hyperplane arrangements, such as the ones giving rise to scattering amplitudes. Let us now flesh out details of this construction in its full generality.

\section{General Formula}

\noindent\textsc{In general}, let us consider a generic arrangement of $k$ hyperplanes on $\mathbb{CP}^m$. They are described with\vspace{-0.5em}
\be
f_i = c_{1i} + \sum_{a=2}^{m+1} c_{ai} \sigma_a,\vspace{-0.5em}
\ee
where $\sigma_a$ for $a=2,3,\ldots,m+1$ are the inhomogeneous coordinates on $\mathbb{CP}^m$. This corresponds to a point in the Grassmannian, $C \in \Gr(m+1,k)$. A given arrangement is non-singular if all maximal minors of $C$ are non-vanishing. The resulting manifold is $X = \mathbb{CP}^m \setminus \cup_{i=1}^{k} \{ f_i \!=\! 0 \}$, and the twist $1$-form $\omega$ is defined as in \eqref{twist}, giving:
\be\label{twist-general}
\omega = \sum_{a=2}^{m+1} \bigg(\! \sum_{i=1}^{k} \frac{\alpha_i\, c_{ai}}{c_{1i}+\sum_{b=2}^{m+1} c_{bi} \sigma_{b}}\bigg)\, d\sigma_a
\ee
with $\sum_{i=1}^{k} \alpha_i = 0$ and $\alpha$'s sufficiently generic.

On this space we introduce the $m$-th twisted cohomology group \cite{aomoto2011theory}:
\be
H^m(X,\nabla_\omega) = \{ \varphi \,\vert\, \nabla_\omega \varphi = 0 \} / \{ \nabla_\omega \xi \},
\ee
where $\nabla_\omega = d + \omega\w$ is the connection and $\xi$ is any smooth $(m-1)$-form on $X$. The dimension of this group is $d = \binom{k-2}{m}$. Its elements are called twisted cocycles. One choice of a basis is the one constructed from cocycles of the form:
\begin{align}
\varphi_L &= d\log \frac{f_{L(1)}}{f_{L(2)}} \w d\log \frac{f_{L(2)}}{f_{L(3)}} \w \cdots \w d\log \frac{f_{L(m)}}{f_{L(m+1)}}\nn\\
&= \widehat{\varphi}_L\, d\sigma_2 \w d\sigma_3 \w \cdots \w d\sigma_{m+1},\label{twisted-cocycles-basis}
\end{align}
for example with $1=L(1)<L(2)<\cdots<L(m+1)<k$ \cite{matsumoto1998}. It is known that an arbitrary twisted cocycle can be expressed in a logarithmic basis \cite{aomoto2011theory}, such as the one above. Similarly, the dual $m$-th twisted cohomology is defined with the connection $\nabla_{-\omega}$, whose basis can be chosen to be the same as in \eqref{twisted-cocycles-basis}.

Intersection numbers are normally computed using the definition \eqref{intro-intersection-number-definition-old} with normalization $1/(2\pi i)^m$ for twisted cocycles $\varphi_L$ and $\varphi_R$ in the original and dual cohomologies respectively, see Appendix~\ref{appendix-a}. In generic arrangements they evaluate, up to an overall sign, to \cite{matsumoto1998}:
\be\label{intersection-number-result}
\la \varphi_L, \varphi_R \ra_\omega =\;\;\;\; \pm\!\!\!\!\!\!\!\!\!\!\!\!\!\! \sum_{\{f_1, f_2, \ldots, f_m\} \in L, R} \frac{1}{\alpha_1\, \alpha_2\, \cdots\, \alpha_m}.
\ee
In singular cases, the above expression requires careful evaluation using blow-ups, see Appendix~\ref{appendix-a}.

Here, we give an alternative formula as an integral localizing on the zeros of $\omega$:
\be\setlength\fboxrule{.5pt}\label{intersection-number-definition}
\color{Maroon}\boxed{\color{black}\la \varphi_L, \varphi_R \ra_\omega = \frac{1}{(-2\pi i)^m} \oint_{\bigwedge_{a=2}^{\!m+1} \{|\omega_a|=\epsilon\}}  \frac{\varphi_L\, \widehat{\varphi}_R}{\prod_{a=2}^{m+1} \omega_a}.}\color{black}
\ee
The above formula is valid even at singular hyperplane arrangements. We used a more precise notation in terms of a multi-dimensional residue around the zeros of \eqref{twist-general}, in place of delta functions localizing the integral like in the example \eqref{intro-intersection-number-definition-new}.

\section{Proof}

\noindent\textsc{Intersection numbers} of twisted cocycles satisfy twisted period relations \cite{cho1995}:
\be\label{twisted-period-relations}
\la \varphi_L, \varphi_R \ra_\omega \!=\! \frac{1}{(2\pi i)^m}\!\!\! \sum_{\alpha,\beta=1}^{d} \!\int_{\mathcal{A}_\alpha}\!\!\!\!\! e^{\int\!\omega} \varphi_L \; \mathbf{H}_{\beta\alpha}^{-1} \int_{\mathcal{B}_\beta} \!\!\!\! e^{-\!\!\int\!\omega} \varphi_R.
\ee
By $\exp\!\int\!\omega$ we denote the multi-valued function $\prod_{i=1}^{k} f_i^{\alpha_i}$ with some choice of a branch. We have two sets of $d$ twisted \emph{cycles} $\{\mathcal{A}_\alpha\}$ and $\{\mathcal{B}_\beta\}$ forming bases of their respective homology groups. Here, $\mathbf{H}$ is the intersection matrix, whose entries are the intersection numbers of these cycles \cite{cho1995,aomoto2011theory,yoshida2013hypergeometric}. Since integrals in the above expression do not generically converge at the same time, they are to be understood in terms of their analytic continuation. In order to use localization arguments, however, we need to define appropriate bases of cycles which fix this problem.

Following Picard--Lefschetz prescription \cite{arnold2012singularities}, we choose bases of twisted cycles to be the paths of steepest descent and ascent of $\exp\!\int\!\omega$ on the same branches, denoted by $\{\mathcal{J}_\alpha\}$ and $\{\mathcal{K}_\beta\}$ respectively. By definition, each of them passes through exactly one critical point $\smash{\sigma_a^{\scriptsize{(\alpha)}}}\!$ of $\exp\!\int\!\omega$, or equivalently a zero of $\omega$. Therefore, cycles intersect only at these points and the intersection matrix becomes an identity matrix, $\mathbf{H}_{\alpha\beta} = \delta_{\alpha\beta}$, giving:
\be
\la \varphi_L, \varphi_R \ra_\omega = \frac{1}{(2\pi i)^m} \sum_{\alpha=1}^{d} \int_{\mathcal{J}_\alpha}\!\!\! e^{\int\!\omega}\, \varphi_L \int_{\mathcal{K}_\alpha} \!\!\! e^{-\!\!\int\!\omega}\, \varphi_R.
\ee
Cycles $\{ \mathcal{K}_\alpha \}$ are now paths of steepest \emph{descent} of $\exp -\!\int\!\omega$. Crucially, this means that each integral on the right-hand side of the above equation converges. Let us rescale $\omega \to \tau \omega$ and take $\tau \to \infty$. In this limit, each integral localizes on precisely one of the $d$ critical points:
\be
\hspace{-2.5em}\lim_{\tau \to \infty}\!\! \la \varphi_L, \varphi_R \ra_{\tau\omega} \!=\! \frac{1}{(-\tau)^m}\! \sum_{\alpha=1}^{d} {\det}^{-1}\! \bigg[\!\, \medmath{\frac{\partial^2 \!\!\int\!\omega}{\partial \sigma_a \partial \sigma_b}} \!\,\bigg] \widehat{\varphi}_L \widehat{\varphi}_R \,\Bigg|_{\sigma_a = \sigma_a^{(\alpha)}}\hspace{-3em},\nn
\ee
where the exponential factors cancel out between the two integrals for each critical point, and $\partial\!\int\!\omega/\partial \sigma_a = \omega_a$ is single-valued. Since the number of critical points equals the dimension of the homology group $d$ \cite{aomoto2011theory}, all zeros of $\omega$ are counted.

On the other hand, intersection numbers $\la \varphi_L, \varphi_R \ra_{\tau\omega}$ are known to scale homogeneously as $\tau^{-m}$, provided that $\varphi_L$ and $\varphi_R$ are expressed in a logarithmic basis \cite{matsumoto1998}, cf. \eqref{intersection-number-result}. We therefore conclude that the above localization formula is exact in $\tau$ and hence we can set $\tau=1$. Expressing the result in terms of a multi-dimensional residue, this proves our claim \eqref{intersection-number-definition}.

\section{Scattering Amplitudes as Intersection Numbers}

\noindent\textsc{Let us now} consider a special case in which the arrangement of hyperplanes produces the moduli space of $n$-punctured Riemann spheres, $X=\mathcal{M}_{0,n}$. The dimension of $X$ is $m=n-3$ and $k=n(n-3)/2+1$ hyperplanes are given by:
\be
\bigcup_{a=2}^{n-2}\; \{\sigma_a = 0\} \;\bigcup_{a=2}^{n-2}\; \{\sigma_a-1 = 0\} \!\!\!\!\!\bigcup_{2\leq a<b \leq n-2}\!\!\!\!\! \{\sigma_a - \sigma_b = 0\}\nn
\ee
with the last one located at infinity. We introduce three coordinates $(\sigma_1, \sigma_{n-1}, \sigma_n) = (0,1,\infty)$ and choose the coefficients $\alpha$ for hyperplanes $\{\sigma_a - \sigma_b = 0 \}$ to be Mandelstam invariants $s_{ab}$. For massless kinematics they add up to zero by momentum conservation. There exists a special kinematic region with all $s_{ab}$ except for $s_{1,n-1}$ being positive \cite{Cachazo:2016ror}, where all $(n-3)!$ zeros of $\omega$ lie in distinct chambers in the real section of the moduli space\footnote{It appears to be true in general that for $\alpha_i > 0,\, i = 1,2,\ldots,k-1$, all zeros of $\omega$ are placed in distinct chambers of $\Re(X)$ not bounded by $\{f_k = 0\}$.}.

The dimension of the cohomology group $d$ undergoes a huge reduction compared to a generic arrangement, from $\binom{n(n-3)/2-1}{n-3}$ to $(n-3)!$ in this singular limit. It also gains an enhanced $\SL(2,\mathbb{C})$ redundancy, $\sigma_a \to (A\sigma_a+B)/(C\sigma_a+D)$ with $AD-BC=1$.

A basis of twisted cocycles can be written using Parke--Taylor forms \cite{Mizera:2017cqs} for the $(n-3)!$ permutations $\alpha$:
\begin{align}
\mathsf{PT}(\alpha) &= d\log {\frac{\sigma_{1,\alpha(2)}}{\sigma_{\alpha(2),\alpha(3)}}} \w \cdots \w d\log {\frac{\sigma_{\alpha(n-3),\alpha(n-2)}}{\sigma_{\alpha(n-2),n-1}}}\nn\\
&= (-1)^n \frac{d\sigma_{\alpha(2)} \w d\sigma_{\alpha(3)} \w \cdots \w d\sigma_{\alpha(n-2)}}{\sigma_{1,\alpha(2)}\, \sigma_{\alpha(2),\alpha(3)}\, \cdots\, \sigma_{\alpha(n-2),n-1}},\label{Parke-Taylor-form}
\end{align}
where $\sigma_{ab} = \sigma_a - \sigma_b$. The twist $1$-form $\omega$ becomes a linear combination of \emph{scattering equations} \cite{Cachazo:2013gna}, $E_a$:
\be\label{twist-scattering-equations}
\omega = \sum_{a=2}^{n-2} \Bigg( \sum_{\substack{b=1\\ b \neq a}}^{n}\frac{s_{ab}}{\sigma_{ab}} \Bigg) d\sigma_a = \sum_{a=2}^{n-2} E_a\, d\sigma_a.\vspace{-0.5em}
\ee

With these assignments, intersection numbers \eqref{intersection-number-definition} become scattering amplitudes in the CHY formulation \cite{Cachazo:2013hca,Cachazo:2013iea,Dolan:2013isa}. Physically, the twist $1$-form \eqref{twist-scattering-equations} translates between singularities of the S-matrix and boundaries of the moduli space. Quantum field theory whose amplitudes are being computed depends on the choice of $\varphi_L$ and $\varphi_R$. For instance, the ingredient $\Pf^\prime \!\Psi$ defined in \cite{Cachazo:2013hca} can be expanded in the basis of twisted cocycles, and the pairings:
\be
\langle \Pf^\prime \!\Psi, \Pf^\prime \!\Psi \rangle_\omega,\quad \langle \textsf{PT}(\alpha), \Pf^\prime \!\Psi \rangle_\omega,\quad \langle \textsf{PT}(\alpha), \mathsf{PT}(\beta) \rangle_\omega,\nn
\ee
give amplitudes of Einstein gravity, Yang--Mills theory, and bi-adjoint scalar respectively.

In this case, equation \eqref{twisted-period-relations} reduces to the so-called \emph{chiral} KLT relation \cite{Kawai:1985xq,Huang:2016bdd,Leite:2016fno}, and $\tau$, which is a rescaling parameter of the Mandelstam invariants, can be identified with the inverse string tension $\alpha'$. In particular, this proves the following two statements:
\begin{enumerate}[label=(\roman*),leftmargin=1.65em]
	\item The result of chiral KLT is a field-theory scattering amplitude in the CHY prescription. This provides mathematical foundations for more physical considerations coming from string theory \cite{Siegel:2015axg,Casali:2016atr,Lee:2017utr,Azevedo:2017yjy,Li:2017emw,Casali:2017mss}.
	\item Since the $\alpha' \to 0$ limit of a closed string amplitude is unaffected up to a sign by the $\alpha' \to -\alpha'$ replacement on one side of the KLT relation, field-theory limit of a closed string amplitude is given by the CHY formula.
\end{enumerate}
Recall that both $\varphi_L$ and $\varphi_R$ are elements of the twisted cohomology groups, and in particular are required to have only logarithmic singularities on the boundaries of the moduli space. Similarly, Mandelstam invariants $s_{ab}$ entering $\omega$ are required to add up to zero, making the above results valid only for massless external states.

\section{Outlook}

\noindent\textsc{Let us put} the results of this letter into a broader perspective. In \cite{Mizera:2017cqs} we found that twisted cycles and cocycles associated to the moduli space $\mathcal{M}_{0,n}$ play a special role in scattering amplitudes. Three types of pairings calculate the following classes of amplitudes:
\begin{center}
\begin{tabular}{ll}
Pairing & Class\\
\hline
$\la \text{\color{black}cocycle}, \text{\color{black}cocycle} \ra$\hspace{1em} & closed string, CHY\\
$\left[ \text{{\color{black}cycle}}, \text{\color{black}cocycle} \ra$ & open string\\
$\left[ \text{{\color{black}cycle}}, \text{{\color{black}cycle}} \right]$ & inverse KLT kernel
\end{tabular}
\end{center}
In this letter we studied intersection numbers of twisted cocycles, which fall into the first class. Within it, the difference between closed string and CHY-type amplitudes comes from a different choice of the dual cohomology group, see \cite{Mizera:2017cqs} and references therein.

Every twisted cocycle has a corresponding cycle, whose boundaries coincide with logarithmic singularities of the former. For instance, Parke--Taylor forms \eqref{Parke-Taylor-form} map to associahedra tiling the moduli space \cite{Devadoss98tessellationsof,Mizera:2017cqs}. Intersection numbers of both cycles and cocycles can then be described using adjacency properties of the associahedra \cite{Mizera:2017cqs,Mizera:2016jhj}, or their linear combinations \cite{CARR20062155,doi:10.1093/imrn/rnn153,Gao:2017dek,Early:2017lku}.

One of the advantages of this way of thinking is that it allows for geometric understanding of relations between different amplitudes, in particular the KLT relations \cite{Kawai:1985xq,Mizera:2017cqs}. They can be summarized using convenient bra-ket notation, see Appendix~\ref{appendix-b}.

It is natural to expect that similar interpretation in terms of intersection numbers can be made at higher loops or for specific theories in four dimensions, especially given recent evidence that field-theory loop integrands can be obtained from genus-zero Riemann surfaces \cite{Geyer:2015bja,Geyer:2016wjx} and obey KLT formulae \cite{He:2016mzd}. The additional challenge is to consider non-generic kinematics on top of singular hyperplane arrangements.

\begin{acknowledgments}
\noindent
We thank Freddy Cachazo, Job Feldbrugge, Alfredo Guevara, Song He, Julio Parra Mart{\'i}nez, Oliver Schlotterer, Piotr Tourkine, and Masaaki Yoshida for useful comments and discussions. This research was supported in part by Perimeter Institute for Theoretical Physics. Research at Perimeter Institute is supported by the Government of Canada through the Department of Innovation, Science and Economic Development Canada and by the Province of Ontario through the Ministry of Research, Innovation and Science.
\end{acknowledgments}

\appendix

\section{\texorpdfstring{}{Appendix A: }One and Two-dimensional Cases}\label{appendix-a}

\noindent\textsc{In order to} contrast the new formula \eqref{intersection-number-definition}, given in the main text, with the standard way of calculating intersection numbers, let us briefly review how the conventional definition \eqref{intro-intersection-number-definition-old} is evaluated in the simplest, one-dimensional, case. Hyperplanes become $k$ points $\{z_i = -c_{1i}/c_{2i}\}$, where $z$ is the inhomogeneous coordinate on $\mathbb{CP}^1$. Logarithmic twisted cocycles $\varphi_L$ can have simple poles only at $z_i$'s. In order to construct $\iota_\omega(\varphi_L)$ with compact support, we must find a cocycle in the same cohomology class which vanishes in a small tubular neighbourhood around each $z_i$. Let us divide the space $X=\mathbb{CP}^1 \setminus \cup_{i=1}^{k} \{z = z_i\}$ into regions as follows:
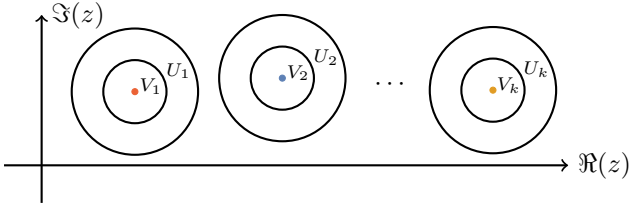
\be
\begin{tikzpicture}[scale=2,baseline=3em]
\draw[thick,->] (-0.25,0) -- (3.5,0) node[right] {$\Re(z)$};
\draw[thick,->] (0,-0.25) -- (0,1.0) node[right] {$\Im(z)$};

\coordinate (p1) at (0.62,0.49);
\draw[color1,fill=color1] (p1) circle (0.02) node[xshift=6,yshift=2] {$\color{black}\scriptstyle{V_1}$};
\draw[thick] (p1) circle (0.21) node[xshift=16.5,yshift=8] {$\scriptstyle{U_1}$};
\draw[thick] (p1) circle (0.42);

\coordinate (p2) at (1.6,0.58);
\draw[color2,fill=color2] (p2) circle (0.02) node[xshift=6,yshift=2] {$\color{black}\scriptstyle{V_2}$};
\draw[thick] (p2) circle (0.21) node[xshift=16.5,yshift=8] {$\scriptstyle{U_2}$};
\draw[thick] (p2) circle (0.42);

\draw (2.33,0.54) node {$\cdots$};

\coordinate (p3) at (3.0,0.50);
\draw[color3,fill=color3] (p3) circle (0.02) node[xshift=6,yshift=2] {$\color{black}\scriptstyle{V_k}$};
\draw[thick] (p3) circle (0.21) node[xshift=16.5,yshift=8] {$\scriptstyle{U_k}$};
\draw[thick] (p3) circle (0.42);
\end{tikzpicture}\nn
\ee

Here $V_i$ and $U_i$ are discs centred at $z_i$ with small radii $0 < \varepsilon_V < \varepsilon_U$. We introduce regulating functions $h_i(z,\bar{z})$ equal to $1$ on $V_i$, $0$ outside of $U_i$ including other $U_{j \neq i}$, and interpolating smoothly in the region $U_i \!\setminus\! V_i$. We then construct:
\begin{align}\label{iota-map}
	\iota_\omega(\varphi_L) &= \varphi_L - \sum_{i=1}^{k} \nabla_{\omega} (h_i \psi_i)\nn\\
	&=\varphi_L - \sum_{i=1}^{k} dh_i \psi_i + h_i \nabla_{\omega} \psi_i,
\end{align}
where $\psi_i$ is the holomorphic function satisfying $\nabla_{\omega} \psi_i = \varphi_L$ on $U_i \setminus \{z_i\}$. It is easy to verify that the unique local solution is given by:
\be\label{psi-residue}
\psi_i = \frac{1}{\alpha_i} \Res_{z=z_i} \{\varphi_L\} + \mathcal{O}(z-z_i).
\ee

From \eqref{iota-map} we see that $\iota_\omega(\varphi_L)$ is in the same cohomology class as $\varphi_L$, but has compact support since $\iota_\omega(\varphi_L) = 0$ on $\cup_{i=1}^{k} V_i$. We also have that $\iota_\omega(\varphi_L) = \varphi_L$ on the outside region $X \setminus \cup_{i=1}^{k} U_i$ and hence vanishes when wedged with $\varphi_R$. The only non-zero contributions to the intersection number come from the annuli $U_i \!\setminus\! V_i$. Using Stokes' theorem we have:\vspace{-.5em}
\begin{align}
	\la \varphi_L, \varphi_R \ra_\omega &= -\frac{1}{2\pi i} \sum_{i=1}^{k} \int_{U_i \setminus V_i} \psi_i\, dh_i \w \varphi_R\nn\\
	&= -\frac{1}{2\pi i} \sum_{i=1}^{k} \int_{U_i \setminus V_i} d(h_i \psi_i \varphi_R)\nn\\
	&= \frac{1}{2\pi i}\sum_{i=1}^{k} \int_{\partial V_i} \psi_i\, \varphi_R.
\end{align}
In the last line we used that $\partial(U_i \!\setminus\! V_i) = \partial U_i - \partial V_i$ and $h_i$ equals $0$ and $1$ on these boundaries respectively. Using \eqref{psi-residue}, the result is a sum over residues of the form $\psi_i \varphi_R$ around $z_i$'s:
\be
\la \varphi_L, \varphi_R \ra_\omega \!=\! \sum_{i=1}^{k} \frac{\Res_{z=z_i} \{\varphi_L\}\! \Res_{z=z_i} \{\varphi_R\}}{\alpha_i} \!= \pm\!\!\!\!\!\! \sum_{\{z_i\} \in L,R} \!\frac{1}{\alpha_i},\nn
\ee
since residues of the twisted cocycles in the basis \eqref{twisted-cocycles-basis} can only be $0$ or $\pm 1$. This is the required result \eqref{intersection-number-result} for $m=1$. A proof of the general formula can be achieved by a generalization of the above derivation to higher dimensions \cite{matsumoto1998}. Alternative proofs were given in \cite{zbMATH03996010,cho1995,Matsumoto1998-2,Ohara98intersectionnumbers,Mimachi2003,Mimachi2004}.

Further generalization of the above formula in the case of singular hyperplane arrangements can be obtained by the use of blow-ups. Let us illustrate it on the two-dimensional case defined with the twist 1-form $\widetilde{\omega}$ from the Introduction. Performing blow-ups near the singular points $(0,0)$ and $(1,1)$ combinatorially yields the configuration:
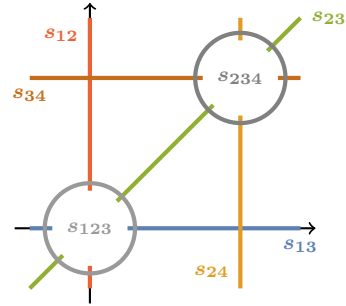
\be
\begin{tikzpicture}[scale=2,baseline=3em]
\draw[thick] (-0.5,0) -- (-0.3,0);
\draw[thick,->] (0.5,0) -- (1.5,0);
\draw[thick] (0,-0.5) -- (0,-0.3);
\draw[thick,->] (0,0.5) -- (0,1.5);
\draw[ultra thick,color1] (0,0.25) -- (0,1.4) node[below left] {$\medmath{\col[1]s_{12} }$};
\draw[ultra thick,color1] (0,-0.4) -- (0,-0.25);
\draw[ultra thick,color2] (0.25,0) -- (1.4,0) node[below] {$\medmath{\col[2]s_{13} }$};
\draw[ultra thick,color2] (-0.4,0) -- (-0.25,0);
\draw[ultra thick,color3] (1,-0.4) node[above left] {$\medmath{\col[3]s_{24} }$} -- (1,0.75);
\draw[ultra thick,color3] (1,1.25) -- (1,1.4);
\draw[ultra thick,color4] (-0.4,1) node[below] {$\medmath{\col[4]s_{34} }$} -- (0.75,1);
\draw[ultra thick,color4] (1.25,1) -- (1.4,1);
\draw[ultra thick,color5] (-0.4,-0.4) -- (-0.18,-0.18);
\draw[ultra thick,color5] (0.18,0.18) -- (0.82,0.82);
\draw[ultra thick,color5] (1.18,1.18) -- (1.4,1.4) node[right] {$\medmath{\col[5]s_{23} }$};
\draw[ultra thick,color7] (0,0) node {$\medmath{\col[7]s_{123} }$} circle (0.3);
\draw[ultra thick,color8] (1,1) node {$\medmath{\col[8]s_{234} }$} circle (0.3);
\end{tikzpicture}\nn
\ee
Here we labelled each hyperplane with its corresponding factor $\alpha$. Blow-ups are illustrated with two effective hypersurfaces, whose $\alpha$ coefficients are sums of those of the three hyperplanes intersecting at these singular points, i.e., ${\col[7]s_{123} } = {\col[1]s_{12} } + {\col[2]s_{13} } + {\col[5]s_{23} }$ and ${\col[8]s_{234} } = {\col[3]s_{24} } + {\col[4]s_{34} } + {\col[5]s_{23} }$. Note that vertices on the opposite sides of the blown-up hypersurfaces are identified.

Similarly, twisted cocycles $\widetilde{\varphi}_L$ and $\widetilde{\varphi}_R$ have to be blown-up. After this procedure, $\widetilde{\varphi}_L$ has logarithmic singularities along hyperplanes corresponding to $({\col[1]s_{12} }, {\col[7]s_{123} }, {\col[5]s_{23} }, {\col[8]s_{234} }, {\col[4]s_{34} } )$, and similarly $\widetilde{\varphi}_R$ gives $({\col[2]s_{13} }, {\col[7]s_{123} }, {\col[5]s_{23} }, {\col[8]s_{234} }, {\col[3]s_{24} } )$. Chambers of the real section of the moduli space bounded by these hyperplanes in both cases correspond to associahedra.

The intersection number is then computed as a sum over all vertices that belong to both blown-up cocycles at the same time, which in this case are two vertices \begin{tikzpicture}[scale=2,baseline=-0.3em]
\draw[ultra thick,color5] (-0.071,-0.071) -- (0.071,0.071);
\draw[ultra thick,color7] (-0.071,0.071) -- (0.071,-0.071);
\end{tikzpicture} and \begin{tikzpicture}[scale=2,baseline=-0.3em]
\draw[ultra thick,color5] (-0.071,-0.071) -- (0.071,0.071);
\draw[ultra thick,color8] (-0.071,0.071) -- (0.071,-0.071);
\end{tikzpicture}, giving:
\be
\left< \widetilde{\varphi}_L, \widetilde{\varphi}_R \right>_{\widetilde{\omega}} = \frac{1}{ \col[5]s_{23} } \left( \frac{1}{ \col[1]s_{12} \!+\! \col[2]s_{13} \!+\! \col[5]s_{23} } + \frac{1}{ \col[3]s_{24} \!+\! \col[4]s_{34} \!+\! \col[5]s_{23} } \right),\nn
\ee
as required.

In general, intersection numbers can be evaluated in the same way by the use of consecutive blow-ups. Combinatorial description of this procedure was given in \cite{Mizera:2017cqs}, where it was shown to arbitrary multiplicity that it leads to bi-adjoint scalar partial amplitudes. The advantage of the new formula for intersection numbers is that it localizes on points far away from singularities of the arrangements, and hence no blow-up procedures are necessary for explicit calculations.

\section{\texorpdfstring{}{Appendix B: }Bra-ket Notation}\label{appendix-b}

\noindent\textsc{Here we introduce} bra-ket notation for twisted cycles and cocycles, which can be used to quickly derive relations between different amplitudes. Consider covariant vectors with a symmetric inner product given by the CHY formula:\vspace{-.5em}
\be\label{inner-product-definition}
\Braket{ \cov[\varphi_L] \! | \! \cov[\varphi_R] } = (-1)^{n-3}\!\! \int \prod_{a=2}^{n-2} d\sigma_a \delta(E_a) \, \widehat{\varphi}_L\, \widehat{\varphi}_R.
\ee
We can choose an $(n-3)!$-dimensional basis spanned by the Parke--Taylor forms \eqref{Parke-Taylor-form}. Their covariant inner product gives a double-partial bi-adjoint scalar amplitude, $m(\alpha | \beta)$:
\be
\Braket{ \! \cov[\PT(\alpha)] \! | \! \cov[\PT(\beta)] \! } = m(\alpha|\beta).
\ee
Using the resolution of identity
\be\label{resolution-of-identity}
\mathds{1} = \sum_{\alpha} \Ket{ \contr[\PT(\alpha)] } \Bra{ \cov[\PT(\alpha)] }
\ee
and employing an implicit summation notation, we find:
\pagebreak

\be \delta^{\alpha}_{\gamma} = \Braket{ \! \contr[\PT(\alpha)] \! | \! \cov[\PT(\gamma)] \! } = \underbrace{\Braket{ \contr[\PT(\alpha)] | \contr[\PT(\beta)] }}_{S_0[\alpha | \beta]} \Braket{ \! \cov[\PT(\beta)] \!|\! \cov[\PT(\gamma)] \! }.\nn
\ee
That is, the contravariant inner product gives a field-theory KLT kernel \cite{BjerrumBohr:2010ta,BjerrumBohr:2010yc,BjerrumBohr:2010hn}, $S_{0}[\alpha|\beta] = m^{-1}(\alpha | \beta)$. Inserting the identity \eqref{resolution-of-identity} into \eqref{inner-product-definition} twice we obtain:
\be
\Braket{ \cov[\varphi_L] \! | \! \cov[\varphi_R] } = \Braket{ \cov[\varphi_L] \! | \! \cov[\PT(\alpha)] \! } \Braket{ \! \contr[\PT(\alpha)] \! | \! \contr[\PT(\beta)] \! } \Braket{ \! \cov[\PT(\beta)] \! | \! \cov[\varphi_R] },\nn
\ee
which is the field-theory KLT relation between different types of amplitudes \cite{Kawai:1985xq,Cachazo:2013iea}. It is also possible to choose an orthonormal basis such that
\be
\Ket{ \contr[\PT(\alpha)] \! } = \delta^{\alpha\beta} \Ket{ \cov[\F(\beta)] \! }, \quad \Braket{ \! \cov[\PT(\alpha)] \! | \! \cov[\F(\beta)] \! } = \delta_{\alpha}^{\beta},
\ee
which can be constructed explicitly with \cite{Mafra:2011nv}:
\be
\widehat{\mathsf{F}}(\beta) = \frac{1}{\sigma_{1,n-1}\sigma_{n-1,n}\sigma_{n,1}}\! \prod_{a=2}^{n-2} \sum_{b=1}^{a-1}\! \frac{s_{\beta(b),\beta(a)}}{\sigma_{\beta(b),\beta(a)}} \frac{\sigma_{\beta(b),n}}{\sigma_{\beta(a),n}}.
\ee

An analogous construction can be made for twisted cycles, $\mathcal{C}$, with square bra-kets and covariant inner products:
\be
\bigg< \cov[\varphi_L] \bigg| \cov[\mathcal{C}_R] \bigg] =\! \int_{\mathcal{C}_R} \!\!\! e^{\int\!\omega} \varphi_L, \quad \bigg[ \cov[\mathcal{C}_L] \bigg| \cov[\varphi_R] \bigg> =\! \int_{\mathcal{C}_L} \!\!\! e^{-\!\!\int\!\omega} \varphi_R.
\ee
Most convenient basis of twisted cycles corresponds to the disk integration regions of open strings, $\C(\alpha)$\footnote{In a special kinematic region \cite{Cachazo:2016ror}, twisted cycles $\{\C(\alpha)\}$ coincide with the Lefschetz thimbles $\{\mathcal{J}_\alpha\}$ \cite{Mizera:2017cqs}, see also \cite{Ohmori:2015sha}.}. The covariant inner product of two square bra-kets is the intersection number of twisted cycles, $m_{\alpha'}(\alpha|\beta)$ \cite{Mizera:2017cqs,Mizera:2016jhj}. Consequently, their contravariant product is the \emph{string theory} KLT kernel, $S_{\alpha'}[\alpha|\beta] = m^{-1}_{\alpha'}(\alpha|\beta)$. In fact, square vectors can be understood as $\alpha'$-deformations of the angle ones \cite{Mizera:2017sen}. In a specific kinematic region \cite{Cachazo:2016ror} we the explicit correspondence between twisted cycles and cocycles is given by: $\mathcal{J}_\alpha \!=\! \C(\alpha) \leftrightarrow \PT(\alpha)$ and $\mathcal{K}_\beta \leftrightarrow \F(\beta)$.

Inserting resolution of identity into \eqref{inner-product-definition} we find:
\be
\Braket{ \cov[\varphi_L] \! | \! \cov[\varphi_R] } = \bigg< \cov[\varphi_L] \bigg| \cov[\C(\alpha)] \bigg] \! \underbrace{\bigg[ \contr[\C(\alpha)] \bigg| \contr[\C(\beta)] \bigg]}_{S_{\alpha^\prime}[\alpha|\beta]} \! \bigg[ \cov[\C(\beta)] \bigg| \cov[\varphi_R] \bigg>,\vspace{-.5em}
\ee
which is the chiral KLT relation \eqref{twisted-period-relations}. A host of other relations between amplitudes can be described using this notation. As mentioned above, analogous statements can be made for the closed string case using a different choice of the dual cohomology group\footnote{Similar idea of relating Koba--Nielsen integrals to certain infinite-dimensional representations of $\SL(2,\mathbb{C})$ dates back to the early years of dual-resonance models \cite{PhysRevD.2.1026,PhysRevD.4.1769,PhysRevD.3.2532,BRINK1972505,Moen1972,Musto1972,Paciello1973}.}.

\bibliography{references}

\begin{thebibliography}{53}%
\makeatletter
\providecommand \@ifxundefined [1]{%
 \@ifx{#1\undefined}
}%
\providecommand \@ifnum [1]{%
 \ifnum #1\expandafter \@firstoftwo
 \else \expandafter \@secondoftwo
 \fi
}%
\providecommand \@ifx [1]{%
 \ifx #1\expandafter \@firstoftwo
 \else \expandafter \@secondoftwo
 \fi
}%
\providecommand \natexlab [1]{#1}%
\providecommand \enquote  [1]{``#1''}%
\providecommand \bibnamefont  [1]{#1}%
\providecommand \bibfnamefont [1]{#1}%
\providecommand \citenamefont [1]{#1}%
\providecommand \href@noop [0]{\@secondoftwo}%
\providecommand \href [0]{\begingroup \@sanitize@url \@href}%
\providecommand \@href[1]{\@@startlink{#1}\@@href}%
\providecommand \@@href[1]{\endgroup#1\@@endlink}%
\providecommand \@sanitize@url [0]{\catcode `\\12\catcode `\$12\catcode
  `\&12\catcode `\#12\catcode `\^12\catcode `\_12\catcode `\%12\relax}%
\providecommand \@@startlink[1]{}%
\providecommand \@@endlink[0]{}%
\providecommand \url  [0]{\begingroup\@sanitize@url \@url }%
\providecommand \@url [1]{\endgroup\@href {#1}{\urlprefix }}%
\providecommand \urlprefix  [0]{URL }%
\providecommand \Eprint [0]{\href }%
\providecommand \doibase [0]{http://dx.doi.org/}%
\providecommand \selectlanguage [0]{\@gobble}%
\providecommand \bibinfo  [0]{\@secondoftwo}%
\providecommand \bibfield  [0]{\@secondoftwo}%
\providecommand \translation [1]{[#1]}%
\providecommand \BibitemOpen [0]{}%
\providecommand \bibitemStop [0]{}%
\providecommand \bibitemNoStop [0]{.\EOS\space}%
\providecommand \EOS [0]{\spacefactor3000\relax}%
\providecommand \BibitemShut  [1]{\csname bibitem#1\endcsname}%
\let\auto@bib@innerbib\@empty
\bibitem [{\citenamefont {Witten}(2004)}]{Witten:2003nn}%
  \BibitemOpen
  \bibfield  {author} {\bibinfo {author} {\bibfnamefont {E.}~\bibnamefont
  {Witten}},\ }\href {\doibase 10.1007/s00220-004-1187-3} {\bibfield  {journal}
  {\bibinfo  {journal} {Commun. Math. Phys.}\ }\textbf {\bibinfo {volume}
  {252}},\ \bibinfo {pages} {189} (\bibinfo {year} {2004})},\ \Eprint
  {http://arxiv.org/abs/hep-th/0312171} {arXiv:hep-th/0312171 [hep-th]}
  \BibitemShut {NoStop}%
\bibitem [{\citenamefont {Arkani-Hamed}\ \emph {et~al.}(2016)\citenamefont
  {Arkani-Hamed}, \citenamefont {Bourjaily}, \citenamefont {Cachazo},
  \citenamefont {Goncharov}, \citenamefont {Postnikov},\ and\ \citenamefont
  {Trnka}}]{ArkaniHamed:2012nw}%
  \BibitemOpen
  \bibfield  {author} {\bibinfo {author} {\bibfnamefont {N.}~\bibnamefont
  {Arkani-Hamed}}, \bibinfo {author} {\bibfnamefont {J.~L.}\ \bibnamefont
  {Bourjaily}}, \bibinfo {author} {\bibfnamefont {F.}~\bibnamefont {Cachazo}},
  \bibinfo {author} {\bibfnamefont {A.~B.}\ \bibnamefont {Goncharov}}, \bibinfo
  {author} {\bibfnamefont {A.}~\bibnamefont {Postnikov}}, \ and\ \bibinfo
  {author} {\bibfnamefont {J.}~\bibnamefont {Trnka}},\ }\href {\doibase
  10.1017/CBO9781316091548} {\emph {\bibinfo {title} {{Scattering Amplitudes
  and the Positive Grassmannian}}}}\ (\bibinfo  {publisher} {Cambridge
  University Press},\ \bibinfo {year} {2016})\ \Eprint
  {http://arxiv.org/abs/1212.5605} {arXiv:1212.5605 [hep-th]} \BibitemShut
  {NoStop}%
\bibitem [{\citenamefont {Arkani-Hamed}\ and\ \citenamefont
  {Trnka}(2014)}]{Arkani-Hamed:2013jha}%
  \BibitemOpen
  \bibfield  {author} {\bibinfo {author} {\bibfnamefont {N.}~\bibnamefont
  {Arkani-Hamed}}\ and\ \bibinfo {author} {\bibfnamefont {J.}~\bibnamefont
  {Trnka}},\ }\href {\doibase 10.1007/JHEP10(2014)030} {\bibfield  {journal}
  {\bibinfo  {journal} {JHEP}\ }\textbf {\bibinfo {volume} {10}},\ \bibinfo
  {pages} {030} (\bibinfo {year} {2014})},\ \Eprint
  {http://arxiv.org/abs/1312.2007} {arXiv:1312.2007 [hep-th]} \BibitemShut
  {NoStop}%
\bibitem [{\citenamefont {Atiyah}\ \emph {et~al.}(2017)\citenamefont {Atiyah},
  \citenamefont {Dunajski},\ and\ \citenamefont {Mason}}]{Atiyah:2017erd}%
  \BibitemOpen
  \bibfield  {author} {\bibinfo {author} {\bibfnamefont {M.}~\bibnamefont
  {Atiyah}}, \bibinfo {author} {\bibfnamefont {M.}~\bibnamefont {Dunajski}}, \
  and\ \bibinfo {author} {\bibfnamefont {L.}~\bibnamefont {Mason}},\ }\href
  {\doibase 10.1098/rspa.2017.0530} {\bibfield  {journal} {\bibinfo  {journal}
  {Proceedings of the Royal Society of London A: Mathematical, Physical and
  Engineering Sciences}\ }\textbf {\bibinfo {volume} {473}} (\bibinfo {year}
  {2017}),\ 10.1098/rspa.2017.0530},\ \Eprint {http://arxiv.org/abs/1704.07464}
  {arXiv:1704.07464 [hep-th]} \BibitemShut {NoStop}%
\bibitem [{\citenamefont {Eisenbud}\ and\ \citenamefont
  {Harris}(2016)}]{eisenbud20163264}%
  \BibitemOpen
  \bibfield  {author} {\bibinfo {author} {\bibfnamefont {D.}~\bibnamefont
  {Eisenbud}}\ and\ \bibinfo {author} {\bibfnamefont {J.}~\bibnamefont
  {Harris}},\ }\href {\doibase 10.1017/CBO9781139062046} {\emph {\bibinfo
  {title} {3264 and all that: A second course in algebraic geometry}}}\
  (\bibinfo  {publisher} {Cambridge University Press},\ \bibinfo {year}
  {2016})\BibitemShut {NoStop}%
\bibitem [{\citenamefont {Cho}\ and\ \citenamefont
  {Matsumoto}(1995)}]{cho1995}%
  \BibitemOpen
  \bibfield  {author} {\bibinfo {author} {\bibfnamefont {K.}~\bibnamefont
  {Cho}}\ and\ \bibinfo {author} {\bibfnamefont {K.}~\bibnamefont
  {Matsumoto}},\ }\href {\doibase 10.1017/S0027763000005304} {\bibfield
  {journal} {\bibinfo  {journal} {Nagoya Math. J.}\ }\textbf {\bibinfo {volume}
  {139}},\ \bibinfo {pages} {67} (\bibinfo {year} {1995})}\BibitemShut
  {NoStop}%
\bibitem [{\citenamefont {Aomoto}\ and\ \citenamefont
  {Kita}(2011)}]{aomoto2011theory}%
  \BibitemOpen
  \bibfield  {author} {\bibinfo {author} {\bibfnamefont {K.}~\bibnamefont
  {Aomoto}}\ and\ \bibinfo {author} {\bibfnamefont {M.}~\bibnamefont {Kita}},\
  }\href {\doibase 10.1007/978-4-431-53938-4} {\emph {\bibinfo {title} {{Theory
  of Hypergeometric Functions}}}},\ Springer Monographs in Mathematics\
  (\bibinfo  {publisher} {Springer Japan},\ \bibinfo {year} {2011})\BibitemShut
  {NoStop}%
\bibitem [{\citenamefont {Yoshida}(2013)}]{yoshida2013hypergeometric}%
  \BibitemOpen
  \bibfield  {author} {\bibinfo {author} {\bibfnamefont {M.}~\bibnamefont
  {Yoshida}},\ }\href {\doibase 10.1007/978-3-322-90166-8} {\emph {\bibinfo
  {title} {{Hypergeometric Functions, My Love: Modular Interpretations of
  Configuration Spaces}}}},\ Aspects of Mathematics\ (\bibinfo  {publisher}
  {Vieweg+Teubner Verlag},\ \bibinfo {year} {2013})\BibitemShut {NoStop}%
\bibitem [{\citenamefont {Kawai}\ \emph {et~al.}(1986)\citenamefont {Kawai},
  \citenamefont {Lewellen},\ and\ \citenamefont {Tye}}]{Kawai:1985xq}%
  \BibitemOpen
  \bibfield  {author} {\bibinfo {author} {\bibfnamefont {H.}~\bibnamefont
  {Kawai}}, \bibinfo {author} {\bibfnamefont {D.~C.}\ \bibnamefont {Lewellen}},
  \ and\ \bibinfo {author} {\bibfnamefont {S.~H.~H.}\ \bibnamefont {Tye}},\
  }\href {\doibase 10.1016/0550-3213(86)90362-7} {\bibfield  {journal}
  {\bibinfo  {journal} {Nucl. Phys.}\ }\textbf {\bibinfo {volume} {B269}},\
  \bibinfo {pages} {1} (\bibinfo {year} {1986})}\BibitemShut {NoStop}%
\bibitem [{\citenamefont {Mizera}(2017{\natexlab{a}})}]{Mizera:2017cqs}%
  \BibitemOpen
  \bibfield  {author} {\bibinfo {author} {\bibfnamefont {S.}~\bibnamefont
  {Mizera}},\ }\href {\doibase 10.1007/JHEP08(2017)097} {\bibfield  {journal}
  {\bibinfo  {journal} {JHEP}\ }\textbf {\bibinfo {volume} {08}},\ \bibinfo
  {pages} {097} (\bibinfo {year} {2017}{\natexlab{a}})},\ \Eprint
  {http://arxiv.org/abs/1706.08527} {arXiv:1706.08527 [hep-th]} \BibitemShut
  {NoStop}%
\bibitem [{\citenamefont {Matsumoto}(1998{\natexlab{a}})}]{matsumoto1998}%
  \BibitemOpen
  \bibfield  {author} {\bibinfo {author} {\bibfnamefont {K.}~\bibnamefont
  {Matsumoto}},\ }\href {https://projecteuclid.org:443/euclid.ojm/1200788347}
  {\bibfield  {journal} {\bibinfo  {journal} {Osaka J. Math.}\ }\textbf
  {\bibinfo {volume} {35}},\ \bibinfo {pages} {873} (\bibinfo {year}
  {1998}{\natexlab{a}})}\BibitemShut {NoStop}%
\bibitem [{\citenamefont {Cachazo}\ \emph
  {et~al.}(2014{\natexlab{a}})\citenamefont {Cachazo}, \citenamefont {He},\
  and\ \citenamefont {Yuan}}]{Cachazo:2013hca}%
  \BibitemOpen
  \bibfield  {author} {\bibinfo {author} {\bibfnamefont {F.}~\bibnamefont
  {Cachazo}}, \bibinfo {author} {\bibfnamefont {S.}~\bibnamefont {He}}, \ and\
  \bibinfo {author} {\bibfnamefont {E.~Y.}\ \bibnamefont {Yuan}},\ }\href
  {\doibase 10.1103/PhysRevLett.113.171601} {\bibfield  {journal} {\bibinfo
  {journal} {Phys. Rev. Lett.}\ }\textbf {\bibinfo {volume} {113}},\ \bibinfo
  {pages} {171601} (\bibinfo {year} {2014}{\natexlab{a}})},\ \Eprint
  {http://arxiv.org/abs/1307.2199} {arXiv:1307.2199 [hep-th]} \BibitemShut
  {NoStop}%
\bibitem [{\citenamefont {Cachazo}\ \emph
  {et~al.}(2014{\natexlab{b}})\citenamefont {Cachazo}, \citenamefont {He},\
  and\ \citenamefont {Yuan}}]{Cachazo:2013iea}%
  \BibitemOpen
  \bibfield  {author} {\bibinfo {author} {\bibfnamefont {F.}~\bibnamefont
  {Cachazo}}, \bibinfo {author} {\bibfnamefont {S.}~\bibnamefont {He}}, \ and\
  \bibinfo {author} {\bibfnamefont {E.~Y.}\ \bibnamefont {Yuan}},\ }\href
  {\doibase 10.1007/JHEP07(2014)033} {\bibfield  {journal} {\bibinfo  {journal}
  {JHEP}\ }\textbf {\bibinfo {volume} {07}},\ \bibinfo {pages} {033} (\bibinfo
  {year} {2014}{\natexlab{b}})},\ \Eprint {http://arxiv.org/abs/1309.0885}
  {arXiv:1309.0885 [hep-th]} \BibitemShut {NoStop}%
\bibitem [{\citenamefont {Bjerrum-Bohr}\ \emph {et~al.}(2012)\citenamefont
  {Bjerrum-Bohr}, \citenamefont {Damgaard}, \citenamefont {Monteiro},\ and\
  \citenamefont {O'Connell}}]{BjerrumBohr:2012mg}%
  \BibitemOpen
  \bibfield  {author} {\bibinfo {author} {\bibfnamefont {N.~E.~J.}\
  \bibnamefont {Bjerrum-Bohr}}, \bibinfo {author} {\bibfnamefont {P.~H.}\
  \bibnamefont {Damgaard}}, \bibinfo {author} {\bibfnamefont {R.}~\bibnamefont
  {Monteiro}}, \ and\ \bibinfo {author} {\bibfnamefont {D.}~\bibnamefont
  {O'Connell}},\ }\href {\doibase 10.1007/JHEP06(2012)061} {\bibfield
  {journal} {\bibinfo  {journal} {JHEP}\ }\textbf {\bibinfo {volume} {06}},\
  \bibinfo {pages} {061} (\bibinfo {year} {2012})},\ \Eprint
  {http://arxiv.org/abs/1203.0944} {arXiv:1203.0944 [hep-th]} \BibitemShut
  {NoStop}%
\bibitem [{\citenamefont {Arnold}\ \emph {et~al.}(2012)\citenamefont {Arnold},
  \citenamefont {Varchenko},\ and\ \citenamefont
  {Gusein-Zade}}]{arnold2012singularities}%
  \BibitemOpen
  \bibfield  {author} {\bibinfo {author} {\bibfnamefont {V.}~\bibnamefont
  {Arnold}}, \bibinfo {author} {\bibfnamefont {A.}~\bibnamefont {Varchenko}}, \
  and\ \bibinfo {author} {\bibfnamefont {S.}~\bibnamefont {Gusein-Zade}},\
  }\href {\doibase 10.1007/978-1-4612-3940-6} {\emph {\bibinfo {title}
  {Singularities of Differentiable Maps: Volume II Monodromy and Asymptotic
  Integrals}}},\ Monographs in Mathematics\ (\bibinfo  {publisher}
  {Birkh{\"a}user Boston},\ \bibinfo {year} {2012})\BibitemShut {NoStop}%
\bibitem [{\citenamefont {Cachazo}\ \emph {et~al.}(2017)\citenamefont
  {Cachazo}, \citenamefont {Mizera},\ and\ \citenamefont
  {Zhang}}]{Cachazo:2016ror}%
  \BibitemOpen
  \bibfield  {author} {\bibinfo {author} {\bibfnamefont {F.}~\bibnamefont
  {Cachazo}}, \bibinfo {author} {\bibfnamefont {S.}~\bibnamefont {Mizera}}, \
  and\ \bibinfo {author} {\bibfnamefont {G.}~\bibnamefont {Zhang}},\ }\href
  {\doibase 10.1007/JHEP03(2017)151} {\bibfield  {journal} {\bibinfo  {journal}
  {JHEP}\ }\textbf {\bibinfo {volume} {03}},\ \bibinfo {pages} {151} (\bibinfo
  {year} {2017})},\ \Eprint {http://arxiv.org/abs/1609.00008} {arXiv:1609.00008
  [hep-th]} \BibitemShut {NoStop}%
\bibitem [{\citenamefont {Cachazo}\ \emph
  {et~al.}(2014{\natexlab{c}})\citenamefont {Cachazo}, \citenamefont {He},\
  and\ \citenamefont {Yuan}}]{Cachazo:2013gna}%
  \BibitemOpen
  \bibfield  {author} {\bibinfo {author} {\bibfnamefont {F.}~\bibnamefont
  {Cachazo}}, \bibinfo {author} {\bibfnamefont {S.}~\bibnamefont {He}}, \ and\
  \bibinfo {author} {\bibfnamefont {E.~Y.}\ \bibnamefont {Yuan}},\ }\href
  {\doibase 10.1103/PhysRevD.90.065001} {\bibfield  {journal} {\bibinfo
  {journal} {Phys. Rev.}\ }\textbf {\bibinfo {volume} {D90}},\ \bibinfo {pages}
  {065001} (\bibinfo {year} {2014}{\natexlab{c}})},\ \Eprint
  {http://arxiv.org/abs/1306.6575} {arXiv:1306.6575 [hep-th]} \BibitemShut
  {NoStop}%
\bibitem [{\citenamefont {Dolan}\ and\ \citenamefont
  {Goddard}(2014)}]{Dolan:2013isa}%
  \BibitemOpen
  \bibfield  {author} {\bibinfo {author} {\bibfnamefont {L.}~\bibnamefont
  {Dolan}}\ and\ \bibinfo {author} {\bibfnamefont {P.}~\bibnamefont
  {Goddard}},\ }\href {\doibase 10.1007/JHEP05(2014)010} {\bibfield  {journal}
  {\bibinfo  {journal} {JHEP}\ }\textbf {\bibinfo {volume} {05}},\ \bibinfo
  {pages} {010} (\bibinfo {year} {2014})},\ \Eprint
  {http://arxiv.org/abs/1311.5200} {arXiv:1311.5200 [hep-th]} \BibitemShut
  {NoStop}%
\bibitem [{\citenamefont {Huang}\ \emph {et~al.}(2016)\citenamefont {Huang},
  \citenamefont {Siegel},\ and\ \citenamefont {Yuan}}]{Huang:2016bdd}%
  \BibitemOpen
  \bibfield  {author} {\bibinfo {author} {\bibfnamefont {Y.-t.}\ \bibnamefont
  {Huang}}, \bibinfo {author} {\bibfnamefont {W.}~\bibnamefont {Siegel}}, \
  and\ \bibinfo {author} {\bibfnamefont {E.~Y.}\ \bibnamefont {Yuan}},\ }\href
  {\doibase 10.1007/JHEP09(2016)101} {\bibfield  {journal} {\bibinfo  {journal}
  {JHEP}\ }\textbf {\bibinfo {volume} {09}},\ \bibinfo {pages} {101} (\bibinfo
  {year} {2016})},\ \Eprint {http://arxiv.org/abs/1603.02588} {arXiv:1603.02588
  [hep-th]} \BibitemShut {NoStop}%
\bibitem [{\citenamefont {Leite}\ and\ \citenamefont
  {Siegel}(2017)}]{Leite:2016fno}%
  \BibitemOpen
  \bibfield  {author} {\bibinfo {author} {\bibfnamefont {M.~M.}\ \bibnamefont
  {Leite}}\ and\ \bibinfo {author} {\bibfnamefont {W.}~\bibnamefont {Siegel}},\
  }\href {\doibase 10.1007/JHEP01(2017)057} {\bibfield  {journal} {\bibinfo
  {journal} {JHEP}\ }\textbf {\bibinfo {volume} {01}},\ \bibinfo {pages} {057}
  (\bibinfo {year} {2017})},\ \Eprint {http://arxiv.org/abs/1610.02052}
  {arXiv:1610.02052 [hep-th]} \BibitemShut {NoStop}%
\bibitem [{\citenamefont {Siegel}(2015)}]{Siegel:2015axg}%
  \BibitemOpen
  \bibfield  {author} {\bibinfo {author} {\bibfnamefont {W.}~\bibnamefont
  {Siegel}},\ }\href@noop {} {\  (\bibinfo {year} {2015})},\ \Eprint
  {http://arxiv.org/abs/1512.02569} {arXiv:1512.02569 [hep-th]} \BibitemShut
  {NoStop}%
\bibitem [{\citenamefont {Casali}\ and\ \citenamefont
  {Tourkine}(2016)}]{Casali:2016atr}%
  \BibitemOpen
  \bibfield  {author} {\bibinfo {author} {\bibfnamefont {E.}~\bibnamefont
  {Casali}}\ and\ \bibinfo {author} {\bibfnamefont {P.}~\bibnamefont
  {Tourkine}},\ }\href {\doibase 10.1007/JHEP11(2016)036} {\bibfield  {journal}
  {\bibinfo  {journal} {JHEP}\ }\textbf {\bibinfo {volume} {11}},\ \bibinfo
  {pages} {036} (\bibinfo {year} {2016})},\ \Eprint
  {http://arxiv.org/abs/1606.05636} {arXiv:1606.05636 [hep-th]} \BibitemShut
  {NoStop}%
\bibitem [{\citenamefont {Lee}\ \emph {et~al.}(2017)\citenamefont {Lee},
  \citenamefont {Rey},\ and\ \citenamefont {Rosabal}}]{Lee:2017utr}%
  \BibitemOpen
  \bibfield  {author} {\bibinfo {author} {\bibfnamefont {K.}~\bibnamefont
  {Lee}}, \bibinfo {author} {\bibfnamefont {S.-J.}\ \bibnamefont {Rey}}, \ and\
  \bibinfo {author} {\bibfnamefont {J.~A.}\ \bibnamefont {Rosabal}},\ }\href
  {\doibase 10.1007/JHEP11(2017)172} {\bibfield  {journal} {\bibinfo  {journal}
  {JHEP}\ }\textbf {\bibinfo {volume} {11}},\ \bibinfo {pages} {172} (\bibinfo
  {year} {2017})},\ \Eprint {http://arxiv.org/abs/1708.05707} {arXiv:1708.05707
  [hep-th]} \BibitemShut {NoStop}%
\bibitem [{\citenamefont {Azevedo}\ and\ \citenamefont
  {Jusinskas}(2017)}]{Azevedo:2017yjy}%
  \BibitemOpen
  \bibfield  {author} {\bibinfo {author} {\bibfnamefont {T.}~\bibnamefont
  {Azevedo}}\ and\ \bibinfo {author} {\bibfnamefont {R.~L.}\ \bibnamefont
  {Jusinskas}},\ }\href {\doibase 10.1007/JHEP10(2017)216} {\bibfield
  {journal} {\bibinfo  {journal} {JHEP}\ }\textbf {\bibinfo {volume} {10}},\
  \bibinfo {pages} {216} (\bibinfo {year} {2017})},\ \Eprint
  {http://arxiv.org/abs/1707.08840} {arXiv:1707.08840 [hep-th]} \BibitemShut
  {NoStop}%
\bibitem [{\citenamefont {Li}\ and\ \citenamefont {Siegel}(2017)}]{Li:2017emw}%
  \BibitemOpen
  \bibfield  {author} {\bibinfo {author} {\bibfnamefont {Y.}~\bibnamefont
  {Li}}\ and\ \bibinfo {author} {\bibfnamefont {W.}~\bibnamefont {Siegel}},\
  }\href@noop {} {\  (\bibinfo {year} {2017})},\ \Eprint
  {http://arxiv.org/abs/1702.07332} {arXiv:1702.07332 [hep-th]} \BibitemShut
  {NoStop}%
\bibitem [{\citenamefont {Casali}\ and\ \citenamefont
  {Tourkine}(2018)}]{Casali:2017mss}%
  \BibitemOpen
  \bibfield  {author} {\bibinfo {author} {\bibfnamefont {E.}~\bibnamefont
  {Casali}}\ and\ \bibinfo {author} {\bibfnamefont {P.}~\bibnamefont
  {Tourkine}},\ }\href {\doibase 10.1103/PhysRevD.97.061902} {\bibfield
  {journal} {\bibinfo  {journal} {Phys. Rev.}\ }\textbf {\bibinfo {volume}
  {D97}},\ \bibinfo {pages} {061902} (\bibinfo {year} {2018})},\ \Eprint
  {http://arxiv.org/abs/1710.01241} {arXiv:1710.01241 [hep-th]} \BibitemShut
  {NoStop}%
\bibitem [{\citenamefont {Devadoss}(1998)}]{Devadoss98tessellationsof}%
  \BibitemOpen
  \bibfield  {author} {\bibinfo {author} {\bibfnamefont {S.~L.}\ \bibnamefont
  {Devadoss}},\ }in\ \href@noop {} {\emph {\bibinfo {booktitle} {Homotopy
  Invariant Algebraic Structures}}}\ (\bibinfo {year} {1998})\ \Eprint
  {http://arxiv.org/abs/math/9807010} {arXiv:math/9807010 [math.AG]}
  \BibitemShut {NoStop}%
\bibitem [{\citenamefont {Mizera}(2017{\natexlab{b}})}]{Mizera:2016jhj}%
  \BibitemOpen
  \bibfield  {author} {\bibinfo {author} {\bibfnamefont {S.}~\bibnamefont
  {Mizera}},\ }\href {\doibase 10.1007/JHEP06(2017)084} {\bibfield  {journal}
  {\bibinfo  {journal} {JHEP}\ }\textbf {\bibinfo {volume} {06}},\ \bibinfo
  {pages} {084} (\bibinfo {year} {2017}{\natexlab{b}})},\ \Eprint
  {http://arxiv.org/abs/1610.04230} {arXiv:1610.04230 [hep-th]} \BibitemShut
  {NoStop}%
\bibitem [{\citenamefont {Carr}\ and\ \citenamefont
  {Devadoss}(2006)}]{CARR20062155}%
  \BibitemOpen
  \bibfield  {author} {\bibinfo {author} {\bibfnamefont {M.}~\bibnamefont
  {Carr}}\ and\ \bibinfo {author} {\bibfnamefont {S.~L.}\ \bibnamefont
  {Devadoss}},\ }\href {\doibase 10.1016/j.topol.2005.08.010} {\bibfield
  {journal} {\bibinfo  {journal} {Topology and its Applications}\ }\textbf
  {\bibinfo {volume} {153}},\ \bibinfo {pages} {2155 } (\bibinfo {year}
  {2006})},\ \Eprint {http://arxiv.org/abs/math/0407229} {arXiv:math/0407229
  [math.QA]} \BibitemShut {NoStop}%
\bibitem [{\citenamefont {Postnikov}(2009)}]{doi:10.1093/imrn/rnn153}%
  \BibitemOpen
  \bibfield  {author} {\bibinfo {author} {\bibfnamefont {A.}~\bibnamefont
  {Postnikov}},\ }\href {\doibase 10.1093/imrn/rnn153} {\bibfield  {journal}
  {\bibinfo  {journal} {International Mathematics Research Notices}\ }\textbf
  {\bibinfo {volume} {2009}},\ \bibinfo {pages} {1026} (\bibinfo {year}
  {2009})},\ \Eprint {http://arxiv.org/abs/math/0507163} {arXiv:math/0507163
  [math.CO]} \BibitemShut {NoStop}%
\bibitem [{\citenamefont {Gao}\ \emph {et~al.}(2017)\citenamefont {Gao},
  \citenamefont {He},\ and\ \citenamefont {Zhang}}]{Gao:2017dek}%
  \BibitemOpen
  \bibfield  {author} {\bibinfo {author} {\bibfnamefont {X.}~\bibnamefont
  {Gao}}, \bibinfo {author} {\bibfnamefont {S.}~\bibnamefont {He}}, \ and\
  \bibinfo {author} {\bibfnamefont {Y.}~\bibnamefont {Zhang}},\ }\href
  {\doibase 10.1007/JHEP11(2017)144} {\bibfield  {journal} {\bibinfo  {journal}
  {JHEP}\ }\textbf {\bibinfo {volume} {11}},\ \bibinfo {pages} {144} (\bibinfo
  {year} {2017})},\ \Eprint {http://arxiv.org/abs/1708.08701} {arXiv:1708.08701
  [hep-th]} \BibitemShut {NoStop}%
\bibitem [{\citenamefont {Early}(2017)}]{Early:2017lku}%
  \BibitemOpen
  \bibfield  {author} {\bibinfo {author} {\bibfnamefont {N.}~\bibnamefont
  {Early}},\ }\href@noop {} {\  (\bibinfo {year} {2017})},\ \Eprint
  {http://arxiv.org/abs/1709.03686} {arXiv:1709.03686 [math.CO]} \BibitemShut
  {NoStop}%
\bibitem [{\citenamefont {Geyer}\ \emph {et~al.}(2015)\citenamefont {Geyer},
  \citenamefont {Mason}, \citenamefont {Monteiro},\ and\ \citenamefont
  {Tourkine}}]{Geyer:2015bja}%
  \BibitemOpen
  \bibfield  {author} {\bibinfo {author} {\bibfnamefont {Y.}~\bibnamefont
  {Geyer}}, \bibinfo {author} {\bibfnamefont {L.}~\bibnamefont {Mason}},
  \bibinfo {author} {\bibfnamefont {R.}~\bibnamefont {Monteiro}}, \ and\
  \bibinfo {author} {\bibfnamefont {P.}~\bibnamefont {Tourkine}},\ }\href
  {\doibase 10.1103/PhysRevLett.115.121603} {\bibfield  {journal} {\bibinfo
  {journal} {Phys. Rev. Lett.}\ }\textbf {\bibinfo {volume} {115}},\ \bibinfo
  {pages} {121603} (\bibinfo {year} {2015})},\ \Eprint
  {http://arxiv.org/abs/1507.00321} {arXiv:1507.00321 [hep-th]} \BibitemShut
  {NoStop}%
\bibitem [{\citenamefont {Geyer}\ \emph {et~al.}(2016)\citenamefont {Geyer},
  \citenamefont {Mason}, \citenamefont {Monteiro},\ and\ \citenamefont
  {Tourkine}}]{Geyer:2016wjx}%
  \BibitemOpen
  \bibfield  {author} {\bibinfo {author} {\bibfnamefont {Y.}~\bibnamefont
  {Geyer}}, \bibinfo {author} {\bibfnamefont {L.}~\bibnamefont {Mason}},
  \bibinfo {author} {\bibfnamefont {R.}~\bibnamefont {Monteiro}}, \ and\
  \bibinfo {author} {\bibfnamefont {P.}~\bibnamefont {Tourkine}},\ }\href
  {\doibase 10.1103/PhysRevD.94.125029} {\bibfield  {journal} {\bibinfo
  {journal} {Phys. Rev.}\ }\textbf {\bibinfo {volume} {D94}},\ \bibinfo {pages}
  {125029} (\bibinfo {year} {2016})},\ \Eprint
  {http://arxiv.org/abs/1607.08887} {arXiv:1607.08887 [hep-th]} \BibitemShut
  {NoStop}%
\bibitem [{\citenamefont {He}\ and\ \citenamefont
  {Schlotterer}(2017)}]{He:2016mzd}%
  \BibitemOpen
  \bibfield  {author} {\bibinfo {author} {\bibfnamefont {S.}~\bibnamefont
  {He}}\ and\ \bibinfo {author} {\bibfnamefont {O.}~\bibnamefont
  {Schlotterer}},\ }\href {\doibase 10.1103/PhysRevLett.118.161601} {\bibfield
  {journal} {\bibinfo  {journal} {Phys. Rev. Lett.}\ }\textbf {\bibinfo
  {volume} {118}},\ \bibinfo {pages} {161601} (\bibinfo {year} {2017})},\
  \Eprint {http://arxiv.org/abs/1612.00417} {arXiv:1612.00417 [hep-th]}
  \BibitemShut {NoStop}%
\bibitem [{\citenamefont {{Deligne}}\ and\ \citenamefont
  {{Mostow}}(1986)}]{zbMATH03996010}%
  \BibitemOpen
  \bibfield  {author} {\bibinfo {author} {\bibfnamefont {P.}~\bibnamefont
  {{Deligne}}}\ and\ \bibinfo {author} {\bibfnamefont {G.}~\bibnamefont
  {{Mostow}}},\ }\href {\doibase 10.1007/BF02831622} {\bibfield  {journal}
  {\bibinfo  {journal} {{Publ. Math., Inst. Hautes \'Etud. Sci.}}\ }\textbf
  {\bibinfo {volume} {63}},\ \bibinfo {pages} {5} (\bibinfo {year}
  {1986})}\BibitemShut {NoStop}%
\bibitem [{\citenamefont {Matsumoto}(1998{\natexlab{b}})}]{Matsumoto1998-2}%
  \BibitemOpen
  \bibfield  {author} {\bibinfo {author} {\bibfnamefont {K.}~\bibnamefont
  {Matsumoto}},\ }\href {http://www.math.kobe-u.ac.jp/~fe/xml/mr1662357.xml}
  {\bibfield  {journal} {\bibinfo  {journal} {Funkcial. Ekvac.}\ }\textbf
  {\bibinfo {volume} {41}},\ \bibinfo {pages} {291} (\bibinfo {year}
  {1998}{\natexlab{b}})}\BibitemShut {NoStop}%
\bibitem [{\citenamefont {Ohara}(1998)}]{Ohara98intersectionnumbers}%
  \BibitemOpen
  \bibfield  {author} {\bibinfo {author} {\bibfnamefont {K.}~\bibnamefont
  {Ohara}},\ }\href {http://www.math.kobe-u.ac.jp/HOME/ohara/Math/980523.ps}
  {\enquote {\bibinfo {title} {{Intersection numbers of twisted cohomology
  groups associated with Selberg-type integrals}},}\ } (\bibinfo {year}
  {1998})\BibitemShut {NoStop}%
\bibitem [{\citenamefont {Mimachi}\ and\ \citenamefont
  {Yoshida}(2003)}]{Mimachi2003}%
  \BibitemOpen
  \bibfield  {author} {\bibinfo {author} {\bibfnamefont {K.}~\bibnamefont
  {Mimachi}}\ and\ \bibinfo {author} {\bibfnamefont {M.}~\bibnamefont
  {Yoshida}},\ }\href {\doibase 10.1007/s00220-002-0766-4} {\bibfield
  {journal} {\bibinfo  {journal} {Communications in Mathematical Physics}\
  }\textbf {\bibinfo {volume} {234}},\ \bibinfo {pages} {339} (\bibinfo {year}
  {2003})}\BibitemShut {NoStop}%
\bibitem [{\citenamefont {Mimachi}\ and\ \citenamefont
  {Yoshida}(2004)}]{Mimachi2004}%
  \BibitemOpen
  \bibfield  {author} {\bibinfo {author} {\bibfnamefont {K.}~\bibnamefont
  {Mimachi}}\ and\ \bibinfo {author} {\bibfnamefont {M.}~\bibnamefont
  {Yoshida}},\ }\href {\doibase 10.1007/s00220-004-1138-z} {\bibfield
  {journal} {\bibinfo  {journal} {Communications in Mathematical Physics}\
  }\textbf {\bibinfo {volume} {250}},\ \bibinfo {pages} {23} (\bibinfo {year}
  {2004})},\ \Eprint {http://arxiv.org/abs/math/0208097} {arXiv:math/0208097
  [math.AG]} \BibitemShut {NoStop}%
\bibitem [{\citenamefont {Bjerrum-Bohr}\ \emph
  {et~al.}(2010{\natexlab{a}})\citenamefont {Bjerrum-Bohr}, \citenamefont
  {Damgaard}, \citenamefont {Feng},\ and\ \citenamefont
  {Sondergaard}}]{BjerrumBohr:2010ta}%
  \BibitemOpen
  \bibfield  {author} {\bibinfo {author} {\bibfnamefont {N.~E.~J.}\
  \bibnamefont {Bjerrum-Bohr}}, \bibinfo {author} {\bibfnamefont {P.~H.}\
  \bibnamefont {Damgaard}}, \bibinfo {author} {\bibfnamefont {B.}~\bibnamefont
  {Feng}}, \ and\ \bibinfo {author} {\bibfnamefont {T.}~\bibnamefont
  {Sondergaard}},\ }\href {\doibase 10.1103/PhysRevD.82.107702} {\bibfield
  {journal} {\bibinfo  {journal} {Phys. Rev.}\ }\textbf {\bibinfo {volume}
  {D82}},\ \bibinfo {pages} {107702} (\bibinfo {year} {2010}{\natexlab{a}})},\
  \Eprint {http://arxiv.org/abs/1005.4367} {arXiv:1005.4367 [hep-th]}
  \BibitemShut {NoStop}%
\bibitem [{\citenamefont {Bjerrum-Bohr}\ \emph
  {et~al.}(2010{\natexlab{b}})\citenamefont {Bjerrum-Bohr}, \citenamefont
  {Damgaard}, \citenamefont {Feng},\ and\ \citenamefont
  {Sondergaard}}]{BjerrumBohr:2010yc}%
  \BibitemOpen
  \bibfield  {author} {\bibinfo {author} {\bibfnamefont {N.~E.~J.}\
  \bibnamefont {Bjerrum-Bohr}}, \bibinfo {author} {\bibfnamefont {P.~H.}\
  \bibnamefont {Damgaard}}, \bibinfo {author} {\bibfnamefont {B.}~\bibnamefont
  {Feng}}, \ and\ \bibinfo {author} {\bibfnamefont {T.}~\bibnamefont
  {Sondergaard}},\ }\href {\doibase 10.1007/JHEP09(2010)067} {\bibfield
  {journal} {\bibinfo  {journal} {JHEP}\ }\textbf {\bibinfo {volume} {09}},\
  \bibinfo {pages} {067} (\bibinfo {year} {2010}{\natexlab{b}})},\ \Eprint
  {http://arxiv.org/abs/1007.3111} {arXiv:1007.3111 [hep-th]} \BibitemShut
  {NoStop}%
\bibitem [{\citenamefont {Bjerrum-Bohr}\ \emph {et~al.}(2011)\citenamefont
  {Bjerrum-Bohr}, \citenamefont {Damgaard}, \citenamefont {Sondergaard},\ and\
  \citenamefont {Vanhove}}]{BjerrumBohr:2010hn}%
  \BibitemOpen
  \bibfield  {author} {\bibinfo {author} {\bibfnamefont {N.~E.~J.}\
  \bibnamefont {Bjerrum-Bohr}}, \bibinfo {author} {\bibfnamefont {P.~H.}\
  \bibnamefont {Damgaard}}, \bibinfo {author} {\bibfnamefont {T.}~\bibnamefont
  {Sondergaard}}, \ and\ \bibinfo {author} {\bibfnamefont {P.}~\bibnamefont
  {Vanhove}},\ }\href {\doibase 10.1007/JHEP01(2011)001} {\bibfield  {journal}
  {\bibinfo  {journal} {JHEP}\ }\textbf {\bibinfo {volume} {01}},\ \bibinfo
  {pages} {001} (\bibinfo {year} {2011})},\ \Eprint
  {http://arxiv.org/abs/1010.3933} {arXiv:1010.3933 [hep-th]} \BibitemShut
  {NoStop}%
\bibitem [{\citenamefont {Mafra}\ \emph {et~al.}(2013)\citenamefont {Mafra},
  \citenamefont {Schlotterer},\ and\ \citenamefont
  {Stieberger}}]{Mafra:2011nv}%
  \BibitemOpen
  \bibfield  {author} {\bibinfo {author} {\bibfnamefont {C.~R.}\ \bibnamefont
  {Mafra}}, \bibinfo {author} {\bibfnamefont {O.}~\bibnamefont {Schlotterer}},
  \ and\ \bibinfo {author} {\bibfnamefont {S.}~\bibnamefont {Stieberger}},\
  }\href {\doibase 10.1016/j.nuclphysb.2013.04.023} {\bibfield  {journal}
  {\bibinfo  {journal} {Nucl. Phys.}\ }\textbf {\bibinfo {volume} {B873}},\
  \bibinfo {pages} {419} (\bibinfo {year} {2013})},\ \Eprint
  {http://arxiv.org/abs/1106.2645} {arXiv:1106.2645 [hep-th]} \BibitemShut
  {NoStop}%
\bibitem [{\citenamefont {Ohmori}(2015)}]{Ohmori:2015sha}%
  \BibitemOpen
  \bibfield  {author} {\bibinfo {author} {\bibfnamefont {K.}~\bibnamefont
  {Ohmori}},\ }\href {\doibase 10.1007/JHEP06(2015)075} {\bibfield  {journal}
  {\bibinfo  {journal} {JHEP}\ }\textbf {\bibinfo {volume} {06}},\ \bibinfo
  {pages} {075} (\bibinfo {year} {2015})},\ \Eprint
  {http://arxiv.org/abs/1504.02675} {arXiv:1504.02675 [hep-th]} \BibitemShut
  {NoStop}%
\bibitem [{\citenamefont {Mizera}\ and\ \citenamefont
  {Zhang}(2017)}]{Mizera:2017sen}%
  \BibitemOpen
  \bibfield  {author} {\bibinfo {author} {\bibfnamefont {S.}~\bibnamefont
  {Mizera}}\ and\ \bibinfo {author} {\bibfnamefont {G.}~\bibnamefont {Zhang}},\
  }\href {\doibase 10.1103/PhysRevD.96.066016} {\bibfield  {journal} {\bibinfo
  {journal} {Phys. Rev.}\ }\textbf {\bibinfo {volume} {D96}},\ \bibinfo {pages}
  {066016} (\bibinfo {year} {2017})},\ \Eprint
  {http://arxiv.org/abs/1705.10323} {arXiv:1705.10323 [hep-th]} \BibitemShut
  {NoStop}%
\bibitem [{\citenamefont {Domokos}\ \emph {et~al.}(1970)\citenamefont
  {Domokos}, \citenamefont {Kovesi-Domokos},\ and\ \citenamefont
  {Schonberg}}]{PhysRevD.2.1026}%
  \BibitemOpen
  \bibfield  {author} {\bibinfo {author} {\bibfnamefont {G.}~\bibnamefont
  {Domokos}}, \bibinfo {author} {\bibfnamefont {S.}~\bibnamefont
  {Kovesi-Domokos}}, \ and\ \bibinfo {author} {\bibfnamefont {E.}~\bibnamefont
  {Schonberg}},\ }\href {\doibase 10.1103/PhysRevD.2.1026} {\bibfield
  {journal} {\bibinfo  {journal} {Phys. Rev. D}\ }\textbf {\bibinfo {volume}
  {2}},\ \bibinfo {pages} {1026} (\bibinfo {year} {1970})}\BibitemShut
  {NoStop}%
\bibitem [{\citenamefont {Bars}\ and\ \citenamefont
  {G\"ursey}(1971)}]{PhysRevD.4.1769}%
  \BibitemOpen
  \bibfield  {author} {\bibinfo {author} {\bibfnamefont {I.}~\bibnamefont
  {Bars}}\ and\ \bibinfo {author} {\bibfnamefont {F.}~\bibnamefont
  {G\"ursey}},\ }\href {\doibase 10.1103/PhysRevD.4.1769} {\bibfield  {journal}
  {\bibinfo  {journal} {Phys. Rev. D}\ }\textbf {\bibinfo {volume} {4}},\
  \bibinfo {pages} {1769} (\bibinfo {year} {1971})}\BibitemShut {NoStop}%
\bibitem [{\citenamefont {Montvay}(1971)}]{PhysRevD.3.2532}%
  \BibitemOpen
  \bibfield  {author} {\bibinfo {author} {\bibfnamefont {I.}~\bibnamefont
  {Montvay}},\ }\href {\doibase 10.1103/PhysRevD.3.2532} {\bibfield  {journal}
  {\bibinfo  {journal} {Phys. Rev. D}\ }\textbf {\bibinfo {volume} {3}},\
  \bibinfo {pages} {2532} (\bibinfo {year} {1971})}\BibitemShut {NoStop}%
\bibitem [{\citenamefont {Brink}\ and\ \citenamefont
  {Kihlberg}(1972)}]{BRINK1972505}%
  \BibitemOpen
  \bibfield  {author} {\bibinfo {author} {\bibfnamefont {L.}~\bibnamefont
  {Brink}}\ and\ \bibinfo {author} {\bibfnamefont {A.}~\bibnamefont
  {Kihlberg}},\ }\href {\doibase 10.1016/0550-3213(72)90081-8} {\bibfield
  {journal} {\bibinfo  {journal} {Nuclear Physics B}\ }\textbf {\bibinfo
  {volume} {46}},\ \bibinfo {pages} {505 } (\bibinfo {year}
  {1972})}\BibitemShut {NoStop}%
\bibitem [{\citenamefont {Moen}(1972)}]{Moen1972}%
  \BibitemOpen
  \bibfield  {author} {\bibinfo {author} {\bibfnamefont {I.~O.}\ \bibnamefont
  {Moen}},\ }\href {\doibase 10.1007/BF02899777} {\bibfield  {journal}
  {\bibinfo  {journal} {Il Nuovo Cimento A (1965-1970)}\ }\textbf {\bibinfo
  {volume} {10}},\ \bibinfo {pages} {784} (\bibinfo {year} {1972})}\BibitemShut
  {NoStop}%
\bibitem [{\citenamefont {Musto}\ \emph {et~al.}(1972)\citenamefont {Musto},
  \citenamefont {Nicodemi}, \citenamefont {Paciello},\ and\ \citenamefont
  {Taglienti}}]{Musto1972}%
  \BibitemOpen
  \bibfield  {author} {\bibinfo {author} {\bibfnamefont {R.}~\bibnamefont
  {Musto}}, \bibinfo {author} {\bibfnamefont {F.}~\bibnamefont {Nicodemi}},
  \bibinfo {author} {\bibfnamefont {M.~L.}\ \bibnamefont {Paciello}}, \ and\
  \bibinfo {author} {\bibfnamefont {B.}~\bibnamefont {Taglienti}},\ }\href
  {\doibase 10.1007/BF02832838} {\bibfield  {journal} {\bibinfo  {journal} {Il
  Nuovo Cimento A (1965-1970)}\ }\textbf {\bibinfo {volume} {7}},\ \bibinfo
  {pages} {407} (\bibinfo {year} {1972})}\BibitemShut {NoStop}%
\bibitem [{\citenamefont {Paciello}\ \emph {et~al.}(1973)\citenamefont
  {Paciello}, \citenamefont {Sciarrino},\ and\ \citenamefont
  {Taglienti}}]{Paciello1973}%
  \BibitemOpen
  \bibfield  {author} {\bibinfo {author} {\bibfnamefont {M.~L.}\ \bibnamefont
  {Paciello}}, \bibinfo {author} {\bibfnamefont {A.}~\bibnamefont {Sciarrino}},
  \ and\ \bibinfo {author} {\bibfnamefont {B.}~\bibnamefont {Taglienti}},\
  }\href {\doibase 10.1007/BF02756276} {\bibfield  {journal} {\bibinfo
  {journal} {Il Nuovo Cimento A (1965-1970)}\ }\textbf {\bibinfo {volume}
  {14}},\ \bibinfo {pages} {591} (\bibinfo {year} {1973})}\BibitemShut
  {NoStop}%
\end{thebibliography}%
	
\end{document}